\documentclass[sn-nature]{sn-jnl}

\usepackage{graphicx}%
\usepackage{multirow}%
\usepackage{amsmath,amssymb,amsfonts}%
\usepackage{amsthm}%
\usepackage{mathrsfs}%
\usepackage[title]{appendix}%
\usepackage{xcolor}%
\usepackage{textcomp}%
\usepackage{manyfoot}%
\usepackage{booktabs}%
\usepackage{algorithm}%
\usepackage{algorithmicx}%
\usepackage{algpseudocode}%
\usepackage{listings}%
\usepackage{verbatim}

\newcommand\mape[1]{{\color{black}#1}}

\newcommand\man[1]{{\color{black}#1}}
\newcommand\usrnec{user NEC}
\newcommand\urlnec{URL NEC}

\begin{document}

\title{Entropy-based detection of Twitter echo chambers}


\author[1,4]{\fnm{Manuel} \sur{Pratelli}}%\email{manuel.pratelli@imtlucca.it}

\author*[2,3,1]{\fnm{Fabio} \sur{Saracco}}\email{fabio.saracco@cref.it}

\author[4,1]{\fnm{Marinella} \sur{Petrocchi}}%\email{marinella.petrocchi@iit.cnr.it}

\affil[1]{%\orgdiv{Department}, 
\orgname{IMT Scuola Alti Studi Lucca}, \orgaddress{\street{Piazza San Francesco 19}, \city{Lucca}, \postcode{55100}, \country{Italy}}}

\affil[2]{\orgname{``Enrico Fermi'' Research Center}, \orgaddress{\street{Via Panisperna 89A}, \city{Rome}, \postcode{00184}, \country{Italy}}}

\affil[3]{\orgname{Institute for Applied Computing ``Mauro Picone'', CNR}, \orgaddress{\street{Via dei Taurini 19}, \city{Rome}, \postcode{00185}, \country{Italy}}}

\affil[4]{\orgname{Istituto di Informatica e Telematica, CNR}, \orgaddress{\street{via G. Moruzzi 1}, \city{Pisa}, \postcode{56124}, \country{Italy}}}


\abstract{Echo chambers, i.e. clusters of users exposed to news \mape{and} opinions in line with their previous beliefs, were observed in many online debates on social platforms.
% Users form an echo chamber when two different phenomena appear at the same time: 1. users interact with others sharing similar opinions; 2. users with similar opinions refer to the same pieces of news. While the concept of echo chambers is clear, there is not a complete agreement about the procedure to infer their presence. Moreover, most of the time the dataset needs an a priori annotation to properly find the opposite factions.\\
%In the present manuscript, 
We propose a completely unbiased entropy-based method for detecting echo chambers. The method is completely agnostic to the nature of the data. In the Italian Twitter debate about the Covid-19 vaccination, we find a limited presence of users in echo chambers (about 0.35\% of all users). 
% due to the limited number of validated users who are exposed to the same news. 
\mape{Nevertheless, their impact on the formation of a common discourse is strong, as users in echo chambers are responsible for nearly a third of the retweets in the original dataset.} \mape{Moreover, in the case study observed, echo chambers appear to be a receptacle for disinformative content.}
%of their discursive communities.
}
\maketitle
\fbox{
\begin{minipage}{12 cm}
\paragraph{Significance statement}
While the concept of echo chambers is clear, a generally accepted method for their detection is still lacking in the literature. Our study provides a general and unbiased method for detecting echo chambers, using entropy-based null models as statistical benchmarks. The rationale is to detect groups of users with similar opinions based on the significant similarity of the main content creators and, similarly, to detect groups of users engaged with the same news. A non-trivial overlap between the two groups indicates the presence of an echo chamber. Using the Italian Twitter debate on the Covid-19 vaccination as a case study, we found that users in echo chambers, while representing a small minority, strongly contribute to the debate, \mape{often disseminating misinformation}.
\end{minipage}
}

\section{Introduction}

% In the virtual world, an {\it echo chamber} is formed when people are exposed only to online information and opinions that align with their existing beliefs and values. 
\mape{In the virtual world, the tendency to seek out information that confirms existing beliefs and to interact with users who share similar opinions leads to the formation of echo chambers, i.e., `bounded, enclosed media spaces that have the potential to both amplify the messages delivered within it and insulate them from rebuttal'~\cite{jamieson08echo,Garrett2009,delvicario2016,Zollo2017Debunking}. A more detailed review of the literature about echo chambers in online social networks can be found in Section 1 of the Supplementary Material.}\\
\mape{We thus have two key events in echo chamber formation: i) interaction between users with similar opinions; ii) exposure of users to the same news articles.} 

\mape{This paper studies the actual presence of echo chambers in social networks by detecting the overlap of the two events. The detection is done by adopting an entropy-based technique.
The platform considered in this study is Twitter/X. From now on, we will refer to the platform by its former name, Twitter, since the analyses were conducted before the change in the company name to X.}

\begin{figure*}[h!]
\begin{center}
\includegraphics[width=.90\textwidth]{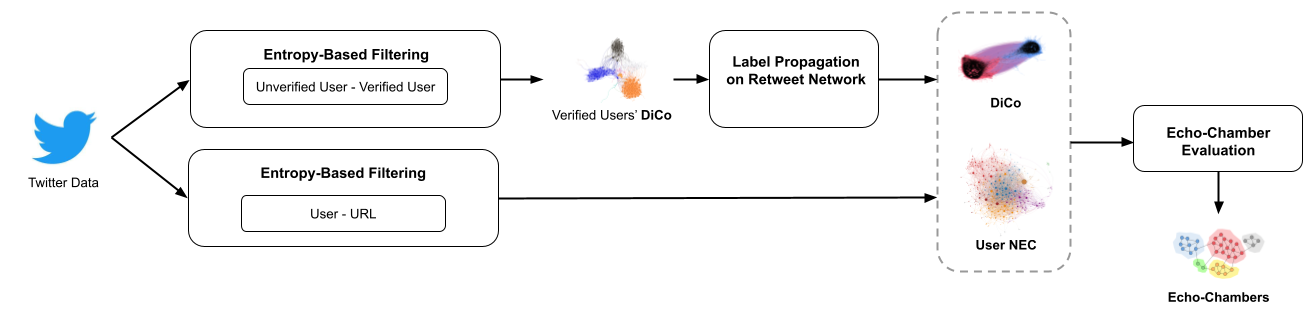}\hfill
\caption{\textbf{Pipeline for Echo-chamber detection.} The upper path focuses on the detection of Discursive Communities (\emph{DiCo}), while the lower one on the detection of News Engagement Communities (\emph{NEC}). Both procedures pass through the statistical validation of empirical data with an entropy-based null model.\label{fig:pipeline}}
\end{center}
\end{figure*}

Recently, entropy-based null models have been introduced in studies of complex networks as an unbiased benchmark capable of revealing non-trivial structures of real systems~\cite{Cimini2018a}, and thus they represent the appropriate framework for our analysis. Fig.~\ref{fig:pipeline} shows how we intend to assess the occurrence of the two events, and, consequently, the occurrence of the echo chamber.

Assessing the opinions of the various accounts is not an easy task, but it can be inferred from the interaction among the various accounts. In Ref.s~\cite{Becatti2019d,Caldarelli2021,Radicioni2021a,Mattei2022} a method to infer the presence of a discursive community, i.e. a group of accounts contributing to the formation of a common discourse, was presented. It is based on verified users, i.e. those accounts for which Twitter has a procedure to check the identity of their owners. Verified accounts mainly belong to politicians, journalists, and celebrities; usually, they are strong creators of contents~\cite{Becatti2019d,Caldarelli2021,Radicioni2021a}. Verified users are among the greatest contributors to the formation of a common discourse. It is possible, then, to let similarities emerge among the content created by verified users, based on the behavior of their common audience in terms of retweets, since retweets are considered a measure of engagement~\cite{Conover2011,Conover2011a,Conover2012}.
In detail, for each pair of verified users, the number of common retweeters is counted. If the number is statistically significant with respect to an entropy-based benchmark, it is validated and we project a link between the verified users' pair. On the monopartite network of verified users thus obtained, we run a community detection algorithm \man{to extract groups of similar verified users (i.e., the \emph{Verified users' DiCo}s in Fig \ref{fig:pipeline})}. Then, the various communities \man{of verified users} are labeled in terms of the users who belong to them (since the users are verified, it is possible, for example, to derive their political leanings and test \emph{a posteriori} the resulting communities).
%provato a spiegare diversamente (lasciato la vecchia version sotto commentata)
%\man{At this point, starting from the retweet network encompassing both verified and unverified users, we utilize a label propagation procedure to extend the labels assigned to the validated verified users even to unverified users.}
%At this point, using a label propagation procedure, the labels are extended to the retweet network of users, for each community, thus obtaining labeled communities of verified and unverified users. 
% merging le 2 versioni
\mape{At this point, the labels are extended to unverified users using a label propagation algorithm~\cite{Raghavan2007a} on the entire retweet network -thus encompassing both verified and unverified users.}

Once again, the use of the retweet network for label propagation is motivated by the fact that there is evidence that users belonging to communities in a retweet network share similar views~\cite{Conover2011,Conover2011a,Conover2012}.  In the following, such communities will be called {\it discursive communities} (or DiCo), and their detection is sketched in the top path of Fig.~\ref{fig:pipeline}. Discursive communities \mape{{\it embrace those users who contribute to the formation of a common discourse}}. %the similarity being captured by retweets to the content created by the same group of verified accounts.\\

Regarding \mape{the exposure to the same news articles}, we approach its assessment by analyzing the ties between the users and the URLs present in their tweets and retweets. 
The bottom path of Fig.~\ref{fig:pipeline} shows the approach. The idea of leveraging the bipartite network of users and URLs was already considered in Ref.~\cite{GuarinoPGC21} for Facebook: in the present case, we translate the idea therein to Twitter. Again, the procedure goes through a comparison between observations and an entropy-based benchmark: if two users tweeted (or retweeted)  the same URLs significantly more than the benchmark, we conclude that the two users share the same information diet in a statistically significant way. 
% Similarly, if two URLs appear in the tweets (or retweets) of the same users significantly more than the benchmark, these URLs pass the validation procedure: 
We can thus 
 identify groups of users sharing the same URLs. 
% and groups of URLs shared by the same users. 
 In the following, user communities 
% and URL communities 
 that passed the validation are called  %respectively, 
 \emph{news engagement communities} of users, 
% and URLs, 
 for short \usrnec s.  \mape{User NECs contextualize the second event: exposure of users to the same news articles.}

 \mape{Now, we were able to identify groups of users exposed to the same news articles (\usrnec ) and groups of users who share a common discourse (DiCo). Users who share a group of the first type and a group of the second type form an echo chamber, provided they interact with each other. The interaction for us is that of retweets since retweets are considered as a form of endorsement to the content created by others~\cite{Conover2011,Conover2011a,Conover2012,Becatti2019d}. Verifying user interactions is an important step because accounts belonging to the same \usrnec \ may either not belong to the same DiCo or, even in the case where they are in the same discursive community, may not interact with each other. In this sense, only users who i) belong to the same \usrnec \ and ii)  belong to the same DiCo and iii) \mape{are connected, even indirectly, through retweets} (i.e., they form a weakly connected component in the retweet network) can be said to represent an echo chamber.}
 %L'interazione per noi e' quella del retweet, since retweets are considered as a form of endorsement to the content created by others~\cite{Conover2011,Conover2011a,Conover2012,Becatti2019d}}

% An echo chamber is then formed when accounts in a \usrnec \ interact among themselves inside the same discursive community. In fact, we can infer the presence of an echo chamber only when there is evidence that users exposed to the same news sources are also exposed to opinions by other users sharing similar viewpoints: the latter is encoded in retweets to users in the same DiCo, since retweets are considered as a form of endorsement to the content created by others~\cite{Conover2011,Conover2011a,Conover2012,Becatti2019d}. 
% Let us remark that such a situation is not given, in principle: accounts belonging to the same \usrnec \ may either not belong to the same DiCo or, even in the case in which are in the same discursive community, not interact among each other. In this sense, only users that i) belong to the same \usrnec \ and ii) also belong to the same DiCo and iii) form a weakly connected component in the retweet network can be claimed to represent an echo chamber.\\

As a case study for evaluating the presence of echo chambers, we consider the online debate on Twitter regarding the Covid-19 vaccination campaign. 
Surprisingly, compared to numerous examples found in the literature, 
we find a limited presence of echo chambers in the analyzed dataset, mainly due to the small dimensions of users' NECs. Although the detected echo chambers are composed of a small number of users with respect to the total number of active users, they play a significant role in terms of retweet interactions, i.e. \mape{the echo chambers that emerged in the case study of the Covid-19 vaccination debate have a significant impact on the creation of a common and cohesive discourse that is not devoid of disinformation. Furthermore,  users who belong to such echo chambers show the same ideas and opinions after years.}

\paragraph{Contributions:} The main contribution of this paper is a novel unbiased method for echo chamber detection. The procedure is based on the very definition of echo chambers and involves the application of an entropy-based null model to discard signals assimilated to noise. 

\paragraph{Research questions:} Keeping in mind that our ultimate goal is to observe if and when discursive communities and news engagement communities of users overlap, thus forming echo chambers, we organize the structure of the paper to answer the following research questions (RQs):

\begin{itemize}
        \item RQ.1: What are the characteristics of the discursive communities (DiCos) and of the news engagement communities of users (users' NECs)?
        %of URLs and 
        % of users  emerging from the case study under investigation?  
        Are there users in common?;%(Sections~\ref{sec:dicos},\ref{sec:necs},\ref{sec:sovrapposizione});
    
        % \item RQ.2: What are the characteristics of the news engagement communities (NEC) 
        % %of URLs and 
        % of users emerging from the case study under investigation? (Section~\ref{sec:necs});
        
        % \item  RQ.3: %Is there an overlap between DiCos and \usrnec? 
        % Users in the same \usrnec\, interact among themselves in the same DiCo? In other words, do validated echo chambers exist in the scenario under investigation? (Section~\ref{sec:sovrapposizione})
        
        \item \mape{RQ.2: What is the relation between the emergent echo chambers and the presence of disinformation, if any? } %(Section~\ref{sec:rep})}
    \end{itemize}

%\todo[Manuel]{Compattare RQ.1 e RQ.2?}

\section{Results}
\subsection{Dataset}\label{sec:datasets}

% \begin{table*}[h!]
% \centering
% \begin{tabular}{l|l}
%     \hline
%    \bf{Keywords} & \bf{English meaning}\\
%     \hline
%     vax, vaccino, vaccini, vaccinarsi &Variants of the word `vaccination' \\
%    novax  & A person against vaccination\\
%     Astrazeneca, Pfizer-BioNTech, Moderna, Sputnik&  Covid-19 vaccines\\
%     greenpass&The certificate of vaccination\\& or of recover from the disease\\
%     \hline
% \end{tabular}
% \caption{\textbf{Keywords used for collecting tweets about the Twitter debate on the Covid-19 vaccination campaign.} Keywords have been searched in Italian, English meanings on the right. \label{tab:keywords}}
% \end{table*}

Our dataset consists of $\sim$1.87M tweets in Italian and $136$k users; nearly $\sim$220k tweets contain URLs. We relied on the Twitter's streaming API and data were collected from September $1^{st}$ to September $24^{th}$ 2021.
The data collection was keyword-based and related to the COVID-19 vaccination online debate. The keywords are compatible with chronicles regarding the vaccination debate in Italy at time of data collection.  We remind the reader that the Twitter's streaming API returns any tweet containing those terms in the text of the tweet, as well as in its metadata.  It is worth noting that it is not always necessary to have each permutation of a specific keyword in the tracking list. For example, the keyword `COVID' would return tweets that contain also both `COVID19' and `COVID-19'. \mape{The keywords for the data collection are in Section 2 of the Supplementary Information.} %Table~\ref{tab:keywords}\\

\subsection{Discursive Communities}\label{sec:dicos}
\begin{figure}
	\centering
	\begin{minipage}{\linewidth}
	     \includegraphics[width=\linewidth]{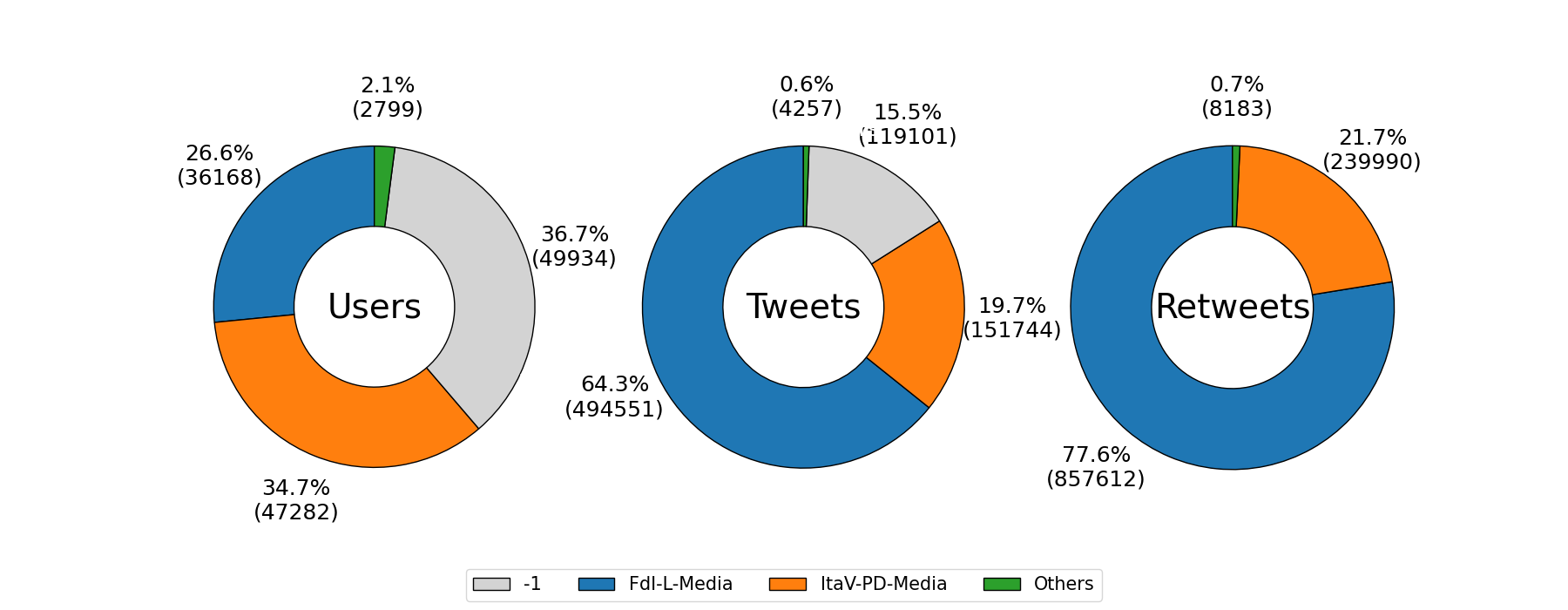}
	      %\subcaption{Blue car}
	      %\label{fig:subfig1}
	\end{minipage}
	% \begin{minipage}{0.45\linewidth}
	%     \includegraphics[width=\linewidth]{image2.jpg}
	%     \subcaption{Speed}
	%     \label{fig:subfig2}
 %        \end{minipage} 
        \\	
	\begin{minipage}{\linewidth}
	    \includegraphics[width=\linewidth]{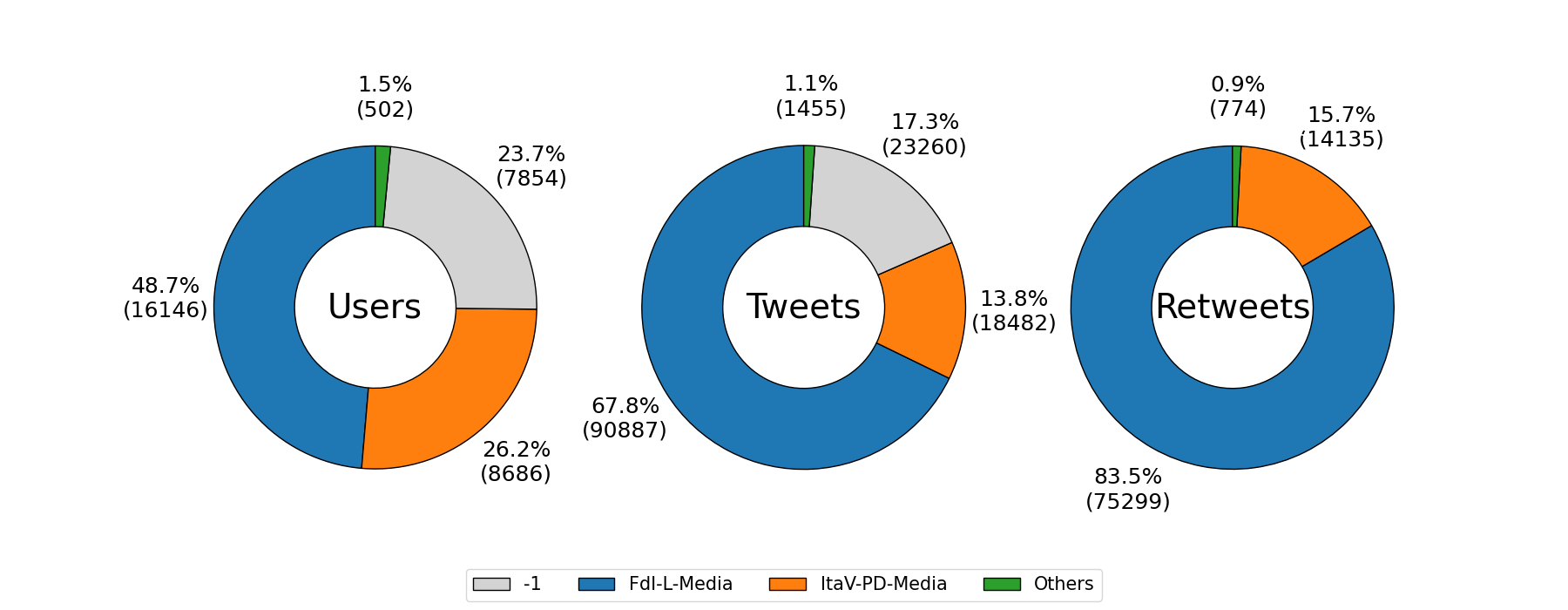}
	 %   \subcaption{Red car}
	%\label{fig:subfig3}
         \end{minipage}
 %         \begin{minipage}{0.45\linewidth}
	%     \includegraphics[width=\linewidth]{image4.jpg}
	%     \subcaption{Performances}
	% \label{fig:subfig4}
 %         \end{minipage}
\caption{Characterization of the main DiCos in terms of the number of users, tweets, and retweets. Charts at the bottom only consider tweets and retweets that contain URLs.\label{fig:dicochar}}
\end{figure}

Fig.~\ref{fig:dicochar} \mape{(top)} describes the characteristics of the main discursive communities (DiCos) that emerge from the data. \mape{We recall that it is possible to assign labels to verified accounts, as the identity of their owner has been certified by the platform. Starting from the original dataset, we run
the community detection algorithm \cite{Blondel2008} on the validated network of verified users and the label propagation algorithm \cite{Raghavan2007a} on the network of retweets of the different communities. In our case study, two main discursive communities emerge, associated with political parties and Italian newspapers. Specifically, 
}
most of the users who are part of a DiCo belong either to the {\sc ItaV-PD-Media} community ($\sim34.7\%$; the community includes journalists and exponents of the Italian parties Italia Viva and Democratic Party) or to the {\sc FdI-L-Media} community ($\sim26.6\%$; the community includes journalists and exponents of the Italian parties Fratelli D'Italia and Lega). About $2.1\%$ of users belong to smaller DiCos, while $\sim36.7\%$ of users do not belong to any DiCo. The {\sc FdI-L-Media} community posted the most new content ($64.3\%$), although it represents about a quarter of all users in our dataset. The {\sc ItaV-PD-Media} community is responsible for $19.7\%$ of the new content, while the remaining $15.5\%$ is posted by users who do not belong to any particular community. In terms of retweets, {\sc FdI-L-Media} is by far the most active community with $77.6\%$ of the retweets.

% \begin{figure*}[h!]
% \includegraphics[width=.99\textwidth]{fig_2.png}\hfill
% \caption{\textbf{Characterization of the main DiCos in terms of the number of users, tweets and retweets.} While the {\sc FdI-L-Media} community is not the most numerous community, its activity in both tweets and retweets abundantly monopolises the debate. \label{fig:donuts_dc}}
% \end{figure*}

% \begin{figure*}
% \centering
% \begin{subfigure}{0.4\textwidth}
%     \includegraphics[width=\textwidth]{fig_2.png}
%     %\caption{Firts subfigure.}
%     \label{fig:donuts_dc}
% \end{subfigure}
% % \hfill
% % \begin{subfigure}{0.4\textwidth}
% %     \includegraphics[width=\textwidth]{example-image}
% %     \caption{Second subfigure.}
% %     \label{fig:second}
% % \end{subfigure}
% \hfill
% \begin{subfigure}{0.4\textwidth}
%     \includegraphics[width=\textwidth]{fig_3.png}
%     %\caption{Third subfigure.}
%     \label{fig:donuts_dc_only_URL}
% \end{subfigure}
        
% \caption{Characterization of the main DiCos in terms of the number of users, tweets and retweets. Donuts at the bottom only consider tweets and retweets that contain URLs.}
% \label{fig:figures}
% \end{figure*}

Fig.~\ref{fig:dicochar} \mape{(bottom)} characterizes DiCos by focusing only on posts containing URLs. In general, the observations made for the top doughnut charts still hold, with the exception that almost half of the users who post tweets with URLs belong to the {\sc FdI-L-Media} community ($48.7\%$). 

% \begin{figure*}[b!]
% \includegraphics[width=.99\textwidth]{fig_3.png}\hfill
% \caption{\textbf{Characterization of the main DiCos, considering only messages including at least one URL}. While the present plots remind the ones in Fig.~\ref{fig:donuts_dc}, here we see that accounts in {\sc FdI-L-Media} dominate also the diffusion of URLs. \label{fig:donuts_dc_only_URL}}
% \end{figure*}

\subsection{News Engagement Communities (NECs) of users}\label{sec:necs}
Table~\ref{tab:usr_community_reliability_compact} shows that of all users who have published at least one post with a URL ($\sim33k$), only 566 are part of a \usrnec, which is less than $2\%$. Accounts in \usrnec s are proportionally much more active in publishing URLs than users not validated by our procedure (67.7 vs. 5.90 URLs per account).

\begin{table}[h!]
\centering
\caption{\textbf{Users in  \usrnec .} Validated users represent a limited minority of all accounts in the debate, i.e. less than 2\% of all users that shared at least a URL. \label{tab:usr_community_reliability_compact}}
\begin{tabular}{lccc}
\hline
Type &  No. Users &  verified &   No. URL \\
\hline
        Non-validated &  32,622 &       434 & 192,334 \\
        Validated &     566 &         1 &   38,345 \\
\hline
\end{tabular}
\end{table}

\begin{figure*}[h!]
\begin{center}
\begin{minipage}{8in}
    \hspace*{.35in}
  \includegraphics[width=.38\textwidth]{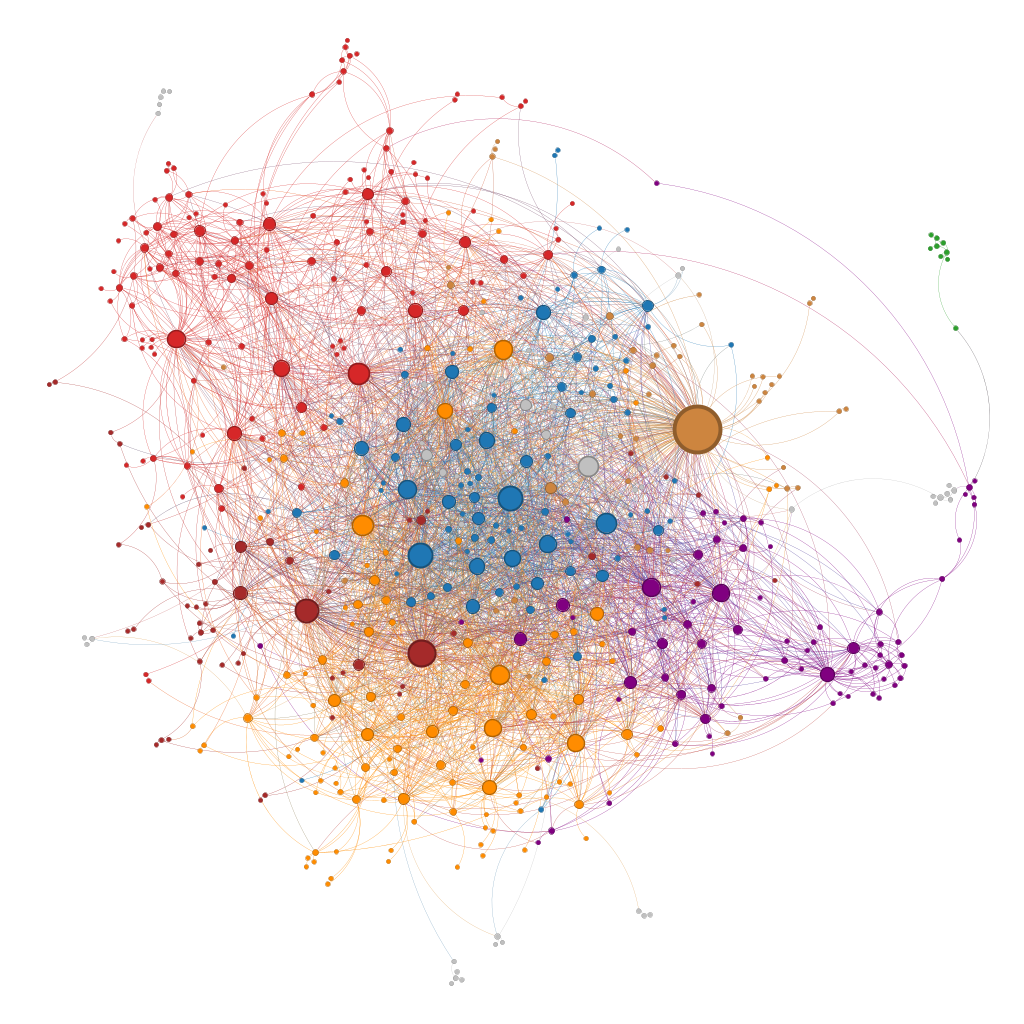}
  \hspace*{.15in}
  \includegraphics[width=.18\textwidth]{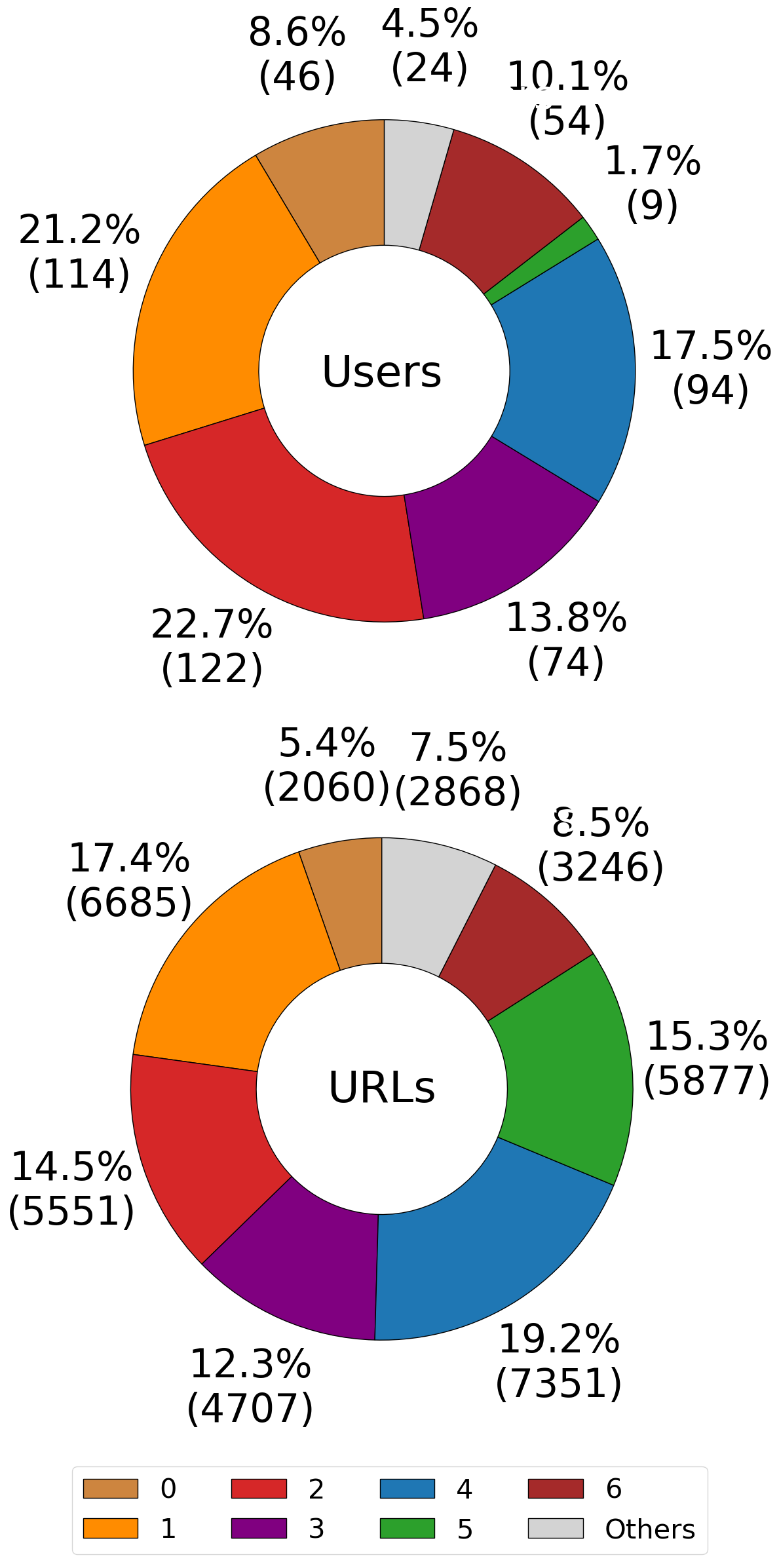}
\end{minipage}

\caption{\textbf{Left: Network representation of \usrnec s. Right, top: percentage (and number) of \usrnec\ users belonging to each group. Right, bottom: Percentage (and number) of URLs disseminated by users belonging to the various \usrnec s.} 
\label{fig:usrec_characterization} }
\end{center}
\end{figure*}

The left panel of Fig.~\ref{fig:usrec_characterization} shows how the 566 users cluster into different \usrnec s, while the right panel provides a statistical view of the 566 users associated with the \usrnec . On the right, the top doughnut chart illustrates the largest communities based on the number of users. Each of these prominent \usrnec s (IDs 0, 1, 2, 3, 4, 5, and 6) accounts for at least $95\%$ of the total user population within this type of community.
Furthermore, the lower doughnut chart shows that these communities have the highest frequency of tweets containing URLs.  Communities 1, 2, 3, 4, and 5 collectively account for over $78\%$ of the total URL traffic generated by all \usrnec\ communities.
\mape{An analogous analysis of the \urlnec s, i.e. the community detected on the validated projection on the layer of the URLs can be found in Section 3 of the Supplementary Information.}

\subsection{Echo chambers}\label{sec:sovrapposizione}

\mape{Our analysis shows that all but 1 of the 566 users in the \usrnec s are also part of the same discursive community, i.e. {\sc FdI-L-Media}. }\mape{This is the discursive community with users affiliated with political parties Fratelli D'Italia and Lega, and news outlets showing similar leanings.}
However, the fact that all users in the \usrnec s belong to the same DiCo only tells us that users with similar `information diets' contribute to the formation of the same discourse, but not that they influence each other and reinforce the opinions of their siblings. In other words, users who refer to the same news sources may never meet on the platform. In fact, the information about who interacts with whom is not used to detect \usrnec s.

As mentioned in the introduction, users in an echo chamber are users who \mape{share a common discourse}, are exposed to the same news sources, and are exposed to the same opinions. Being exposed to the same opinions, translated to Twitter, means that they retweet each other. 
In this sense, if users in the same \usrnec \ form a (weakly) connected component in the same DiCo-induced subgraph of the retweet network (i.e., if there is a flow of influence in the retweet network that is restricted to nodes in the same discursive community), they form an echo chamber.

\mape{The analysis of the weakly connected component shows that 92 users do not belong to it. This leaves 473 users trapped in echo chambers.} In particular, all users in \usrnec s 8, 9, and 10 did not retweet others in the same \usrnec\ on the topic under analysis. Regarding the other \usrnec s, we observe that for each of them, most of the nodes form echo chambers. 
In the following, echo chambers inherit the ID of their \usrnec. 
Some echo chambers are relatively large: for example, those induced by \usrnec s 1 and 2 contain more than 100 nodes.

To study how much users in echo chambers are connected, we use the undirected clustering coefficient: ignoring the direction of the edges, it captures the observed frequency of interactions between the neighbors of each node~\cite{Caldarelli2010}. %\fab{We do not consider the direction of the interactions, since it is not particularly relevant at this point who is retweeting whom}. 
% In the following, we compute the clustering coefficients for the different \usrnec s on the subgraph of the retweet network that contains only nodes in the same \usrnec.

\begin{figure*}[b!]
\includegraphics[width=.90\textwidth]{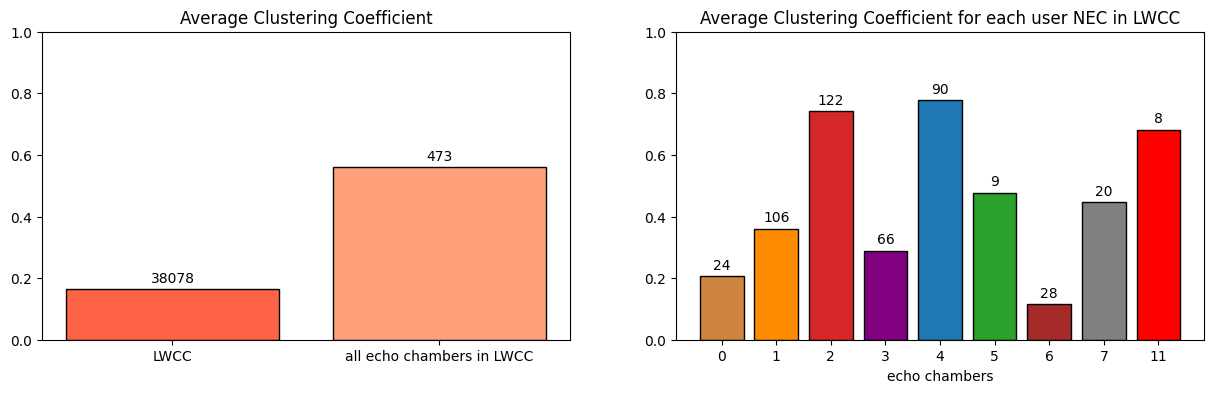}\hfill
\caption{\textbf{Left: average clustering coefficient \mape{measured on the LWCC of the retweet network restricted to users of {\sc FdI-L-Media}} 
%for LWCC of the subgraph of the retweet network induced by {\sc FdI-L-Media} 
and \mape{measured on} all users belonging to echo chambers. Right: average clustering coefficient calculated on each echo chamber.} Each echo chamber inherits the ID and the color from its \usrnec. The number of users in the echo chamber is shown at the top of each bar. 
\label{img:usrnec_clustering_coefficients}}
\end{figure*}

We compare the clustering coefficients of the echo chambers with the one measured on the Largest Weakly Connected Component (LWCC) of the retweet network restricted to users in the {\sc FdI-L-Media} DiCo. In this way, we have a benchmark that captures the main contribution to the discourse to which the echo chambers belong. 
The clustering coefficient associated with users in echo chambers is more than three times as high as that for other users within the LWCC ($0.56$ compared to $0.16$,  left panel of Fig.~\ref{img:usrnec_clustering_coefficients}). 
We then examine the average clustering coefficient within each echo chamber.  The right panel of Fig.~\ref{img:usrnec_clustering_coefficients} shows that the average clustering coefficients of echo chambers $2$, $4$, and $11$ are greater than 0.6.

\mape{
High values of the clustering coefficient imply that accounts are highly connected and frequently retweet each other. Therefore, we can conclude that their endorsement activity contributes to the reinforcement of their opinions. 
%
% High values of the clustering coefficient imply that the activity of users in these groups contributes to the reinforcement of their opinions since all accounts therein are particularly connected among themselves and frequently endorse the content shared by their friends in the \usrnec. 
Such a conclusion is confirmed by a manual examination of the content shared by users in echo chambers after almost 2 years. At the time of the data collection, the opinions of the users were strongly against the Covid-19 vaccination. After 2 years, the positions of the users there still adhere to conspiracy theories and have become particularly extreme. }
%More details can be found in the Supplementary Information.

% As shown in Fig.~\ref{img:USR_per_comm_purity_T_N}, among the greatest echo chambers, 4 has the greatest tendency to share low credibility content. The target of our analysis is to check if the topics and narrative captured by our dataset are still present in the present narrative of the same users.  

\mape{To provide a concrete example, we will focus on the content shared in echo chamber 4.
In practice, we first manually extract the main narratives from the news shared within echo chamber 4, focusing on the users with the highest number of followers at the time of data collection. Then, still focusing on the users with the highest number of followers, we analyze whether there are signals of these narratives in their most recent posts (as of June 7, 2023) and which narratives they currently support.
In echo chamber 4, there are about 1.7k unique news that have been shared about 7.3k times in total. First, we exclude the news with connection errors at the time of this analysis ($1$k shares) and those that have been shared less than 10 times. Then, we analyze the resulting news narratives, which amount to $\sim3.3$k shares and 146 unique URLs from 51 different domains. By classifying only these 146 news stories,  
%(over 1.7k), 
we cover about $\sim45\%$ of the total URL traffic within echo chamber 4. Table~\ref{tab:usr_nec_supported_narratives} shows the narratives' distribution and their descriptions: the main 8 narratives are all against vaccination and government regulations.}

%\begin{figure}[h!]\begin{center}
%\includegraphics[width=.40\textwidth]{fig_SI_4.png}\hfill
%%\caption{\textbf{\mape{Narratives in echo chamber 4.}}
%\label{img:narrative_distribution}}
%\end{center}
%\end{figure}

\begin{table}[ht]
\centering
%\begin{center}
\caption{\textbf{Narratives' descriptions. Echo chamber 4}. 
\label{tab:usr_nec_supported_narratives}
}
\begin{tabular}{lll}
\hline

\bf{ID} & \bf{Narrative} & \bf{\#url supporting the Narrative}\\
\hline
1 & Protests against the Covid-19 green pass & 673\\
%& il greenpass è discriminatorio; proteste di persone \\
%& & o gruppi  (professori, aziende) contro greenpass \\
2 & Comments on presumed deaths after vaccine & 536\\
% Several cases of people who died because of the vaccination\\ 
% & (aneurysm, meningitis, \dots).\\
3 & comments on presumed injuries after vaccine & 510\\
% & (bleeding, myocarditis, pericarditis, neurological problems, excruciating pain)\\
4 & Statements made by politicians against vaccinations & 495\\
5& Against mandatory vaccination & 441\\
%5 & The obligation cannot be imposed since it is unconstitutional.\\
& News about VIPs rejecting the vaccine. & \\
& Ineffectiveness of vaccines. & \\
6 & Mattarella incites to hatred no-vax. Experts reject the third dose & 308\\
7 & (Manipulated) data about vaccine hazard versus efficacy & 196\\ 
& and hospitalizations or infections despite vaccination & \\
8 & COVID-19 vaccines are still too experimental  & 126\\
&  Police forces were not vaccinated. Support to views of no-vax doctors.\\ 
&VIPs and high-ups pretend to be vaccinated, but actually are not & \\ 
%due to the known dangers of vaccinations.\\
\hline
\end{tabular}
%\end{center}
\end{table}
%\end{center}

%\mape{As shown in Table~\ref{tab:usr_nec_supported_narratives}} the main 8 narratives are all against vaccination and government regulations. 
% bringing up the ineffectiveness, if not harmfulness, of the immunization procedure, as well as conspiracy theories. The details are in Table~\ref{tab:usr_nec_supported_narratives}.\\

% Subsequently, our attention shifts towards users in echo chamber 4, who are presumed to hold greater influence due to their substantial number of followers.\\ 
% We find that the main narratives receive substantial support from the tweets of these influential users within this subset, even those not including URLs. Such an outcome suggests that it is indeed feasible to characterize the primary narrative prevalent within an echo chamber by examining the shared news, which proves to be a relatively straightforward task.\\

\mape{Table~\ref{tab:usr_nec_top_10_followers_users_narratives} shows the narratives supported by the users in echo chamber 4 with the most followers, almost two years after data collection (June 7, 2023).
% Regarding the main narrative recently endorsed by the most influential users in echo chamber 4, the results are presented in Table \ref{tab:usr_nec_top_10_followers_users_narratives}. Users under analysis
Users hold extreme views on current controversial issues such as the war in Ukraine, migrants, and LGBT issues. Remarkably, conspiracy theories about vaccines are still present in their narratives.}

%\begin{itemize}
%    \item Future work: abbaimo qui considerato due tipi di similarità ma niente vieta di utilizzarne anche altre (hashtags, mentions, \dots)
%\end{itemize}

\begin{table}[ht]
\centering
\caption{{\bf Main narratives supported in recent posts (as of June 7, 2023) by users in echo chamber 4 with the most followers.} Users are anonymized.}
\label{tab:usr_nec_top_10_followers_users_narratives}
%\begin{center}
\begin{tabular}{cc|l}
\hline
\bf{User} & \bf{Followers} & \bf{Supported Narratives}\\
\hline
user\_1		& 36926 & no-migrants, no-vax, anti-EU\\% (insetti) \\
user\_2	& 6929 & pro-Russia, no-vax, no-LGBT\\
user\_3	& 6335 & pro-Russia, no-migrants, anti-EU, conspiracy theories\\
user\_4	& 4164 & no-vax, no-migrants, pro-Russian\\
user\_5		& 3117 & no-vax\\
user\_6		& 2668 & pro-Russian, against the Italian government, no-vax\\
user\_7	& 2641 & suspended\\
user\_8	& 2448 & conspiracy theories, no-vax, no-LGBT, against the Italian government, anti-EU \\%(insetti)\\
user\_9		& 2355 & religious posts, no-green pass, no-vax\\
user\_10	& 2316 & against Italian government, no-vax\\

\hline
\end{tabular}
%\end{center}
\end{table}

\begin{figure}[h!]
\begin{center}
\includegraphics[width=.7\textwidth]{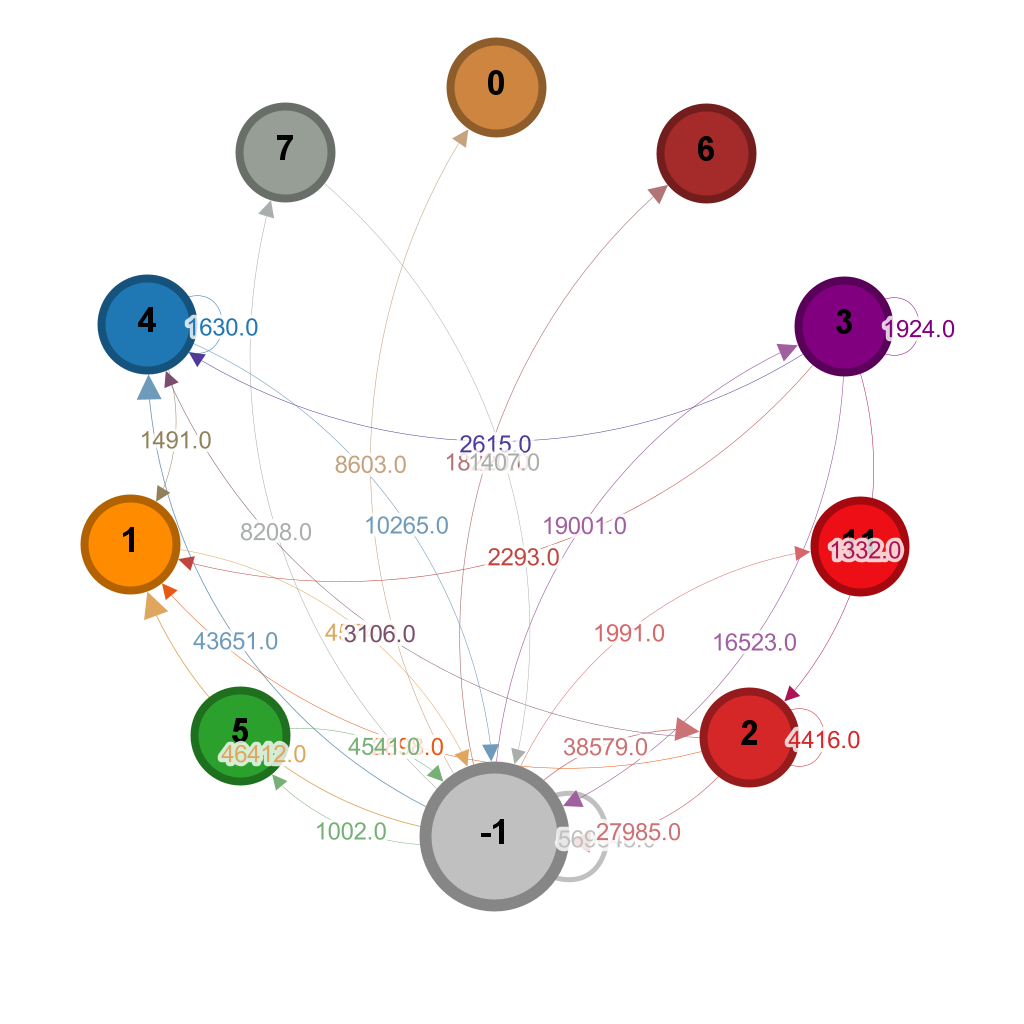}
\end{center}
\caption{\textbf{Retweet network for {\sc FdI-L-Media} DiCo, aggregated with respect to echo chambers.} Node $-1$ represents users who do not belong to an echo chamber.  Edges indicate the number of retweets between different user groups; weights less than $1k$ have been filtered out.
\label{img:usrnec_aggregated_by_type}}
\end{figure}

\subsection{Echo chambers, \mape{their role in the common discourse and the plague of misinformation} }
 
\mape{Figure~\ref{img:usrnec_aggregated_by_type} shows the flow of retweets within an echo chamber and between different echo chambers.  }

Node $-1$ represents all nodes in the DiCo that are not part of an echo chamber, and an arrow indicates that tweets published by the source group are retweeted by a certain number of users in the target group. Self-loops represent retweet activity within the same group. The values on the edges indicate the number of retweets associated with each interaction.
Although the echo chambers are composed of a small number of users (on the order of $10^2$, compared to the total number of DiCo users, on the order of $10^4$), they contribute significantly to the DiCo's retweet activity. Echo chambers are involved in generating about 288k retweets, while users not in echo chambers generate about 569k retweets. More specifically, echo chambers 2 and 3 are mainly composed of popular users (in terms of received retweets), while others are mainly composed of retweeting users ($0$, $1$, $4$).

\mape{To quantify the presence of misinformation in echo chambers, we have tagged URLs in our dataset that point to news sites. The labels are those that the NewsGuard journalistic organization has assigned to online media outlets\footnote{https://www.newsguardtech.com/solutions/newsguard/}. Use of the labels has been licensed to the authors of this article. More details about the reputability measure implemented in the present manuscript can be found in Section 4 of the Supplementary Information.}

% For each type of community (i.e., DiCo, \usrnec s and echo chambers), we label all messages including URLs pointing to news articles according to the annotation of the domain of news publishers as we did in Subsection \ref{sec:NECURL}. Let us remark that if the same URL is shared multiple times by users in the same group, this multiplicity is accounted for in the present analysis.

\begin{figure}[h!]
\begin{center}
\includegraphics[width=.7\textwidth]{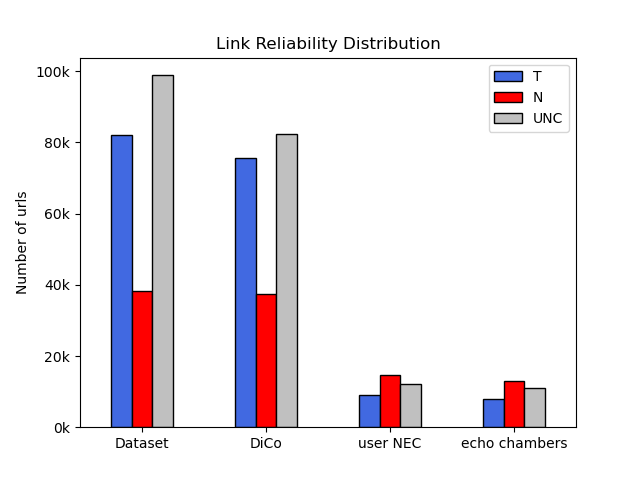}
\caption{\textbf{\mape{Number of distinct URLs pointing to  news publishers tagged as `Trustworthy' (T), `Not trustworthy' (N), or `Unclassified' (UNC) for the entire dataset and for each type of users' community (DiCos, \usrnec s, echo chambers.)}}
\label{fig:frequence_T_N_all}}
\end{center}
\end{figure}

\mape{Figure \ref{fig:frequence_T_N_all} shows the number of URLs pointing to news from publishers that NewsGuard classifies as `Trustworthy' (T), `Not Trustworthy' (N), and `Unclassified' (UNC) for the entire dataset and for each type of user community. If the same URL is shared multiple times by users in the same group, this multiplicity is taken into account in the analysis.

% the overall dataset traffic that a particular community covers for a certain nutrition label provided by NewsGuard. Notably, 
The first observation is that the differences between \usrnec s and echo chambers are negligible. Second, DiCos cover almost the entire volume of both T and N traffic. 
Remarkably, while the ratio between untrusted and trusted URLs is around 0.5 for the entire dataset, the ratio is almost reversed for echo chambers: the frequency of N news sources is almost twice that of T news sources.}

\mape{Section 5 of the Supplementary Information will show even more alarming results regarding the spread of disinformation in echo chambers. We do not show these results in the main text due to space limitations, but i) the probability that a link shared by a user in an echo chamber refers to an untrustworthy news source is 0.377, compared to 0.129 for users outside echo chambers; ii) the probability that a link shared by a user in an echo chamber refers to a trustworthy news source is 0.232, compared to 0.379 for users outside echo chambers. }

% Remarkably, in our dataset users in echo chambers are particularly affected by misinformation: the probability that a link shared by an echo chamber users refers to a not-trustworthy source is 0.377, against 0.129 for other users. The trend is reverted when considering the trustworthy sources: the probability that a link shared by a user in an echo chamber refer to a trustworthy source is 0.232, against 0.379 for the other users. More information regarding the quality of the content shared in the echo chambers can be found in Section 4 of Supplementary Information.

\section{Conclusion}\label{sec:discussion}
In this paper, we propose a novel unbiased method to detect echo chambers. The method is mainly based on two observations. 
First, echo chambers form when users interact with others who share similar opinions and refer to the same news.
% In fact, placement in a social context with the same political orientation reinforces users' beliefs and leads to polarization. 
Second, a proper null model should be implemented to detect a true signal. This necessity has recently been highlighted in the literature on online social networks and has been shown to be particularly important for the detection of non-trivial phenomena~\cite{Becatti2019d,Caldarelli2020b, Caldarelli2021,Mattei2022,declerck2022a,declerck2022b}. 
Our echo chamber detection method is based on the validation of observed structures by comparison with a proper maximum entropy null model; the maximization of entropy guarantees the unbiased nature of the benchmark.

We tested our procedure on a dataset containing the Italian Twitter debate on Covid-19 vaccination: we found that our procedure detects a low presence of echo chambers (just under 0.35\% of all users in our dataset belong to an echo chamber).   All the echo chambers we detected are part of the same discursive community, i.e. a community of users with similar political positions. Even if their dimension in terms of the number of users is limited, their impact on the shared discourse is remarkable: echo chambers are responsible for almost a third of the retweets in their discursive communities.
%Furthermore, in most of the echo chambers, the focus is on unreliable news sources, as certified by the fact-checker NewsGuard (\url{https://www.newsguardtech.com}).\\

The methodology can be extended to other online social networks. In fact, it is based on i) the analysis of the activity of accounts that share URLs to news sources and ii) the detection of discursive communities. While the extension of the former to other online social networks is straightforward, the latter may be more problematic: in the present case, we used the activity of verified users, who are among the main content creators in Twitter~\cite{Becatti2019d}, but not all social platforms have such certification. Nevertheless, when analyzing other platforms, we can still focus on users who are particularly active in creating new content, such as influential users as defined in~\cite{Gonzalez2013}.

Not unlike other studies, our study has some limitations, which we believe do not affect our final conclusions.
First, it may be argued that the validation procedure is quite strict: the validation of multiple p-values leads to the validation of extreme events. While this is true, it is the only way to eliminate random noise from the system and analyze the true signal \mape{(see Sections 7 in the Supplementary Information for more details)}. 
Finally, the main idea of echo chambers is that users follow accounts with similar ideas, while in the present study only the retweet network is used, not the information about friendships.
Still, the retweet network captures the effective interactions with interesting content as perceived by different users, whether it comes from friends or is suggested by the platform itself: focusing only on friendship will not fully capture the effect of the platform's recommendation algorithm.
%Finally, another criticism could be about the news engagement: in fact, we are considering users sharing URLs, i.e. those that not only liked the content but decided to share it either via tweets or retweets. Of course, such an approach has an intrinsic limitation: by capturing only tweets and retweets, we lost part of the information about users being exposed to a piece of news, but not engaged enough to share it via the online social network platform. %On the other hand, by focusing only on the URLs shared, we missed the polarising content that does not display any URL. While the former is sort of unavoidable, the latter surely represents a conservative approach. However, please consider also that by focusing on URLs, we limit the \emph{a priori} intervention of researchers: all our annotations were made by a third-party fact-checker (i.e. NewsGuard, \url{https://www.newsguardtech.com}). In the analysis of texts, the authors should annotate various keywords that they assign to different political leanings, thus introducing a bias in the study, due to the authors' perception. 

\section{Methods}\label{sec:methods}
\subsection{Network analysis methods}
Recently, De Clerck {\it et al.} stressed the importance of using proper statistical benchmarks for the analyses of Online Social Networks~\cite{declerck2022a,declerck2022b}: in fact, such systems are affected by strong noise and detecting genuine signals is fundamental in order to drive the proper conclusions.  In fact, our procedure for the detection of echo chambers is based on the statistical validation of different co-occurrence networks. %, i.e. 
Co-occurrences are implicitly based on a bipartite structure: if we count, for instance, the number of URLs that have been shared by both a pair of users, we are implicitly projecting the information contained in a bipartite network in which layers represent users and URLs on the layer of users. Therefore, including the bipartite information in the analysis of the observed co-occurrences provides a more accurate benchmark.

A general framework for providing unbiased benchmarks for the analysis of complex networks was recently proposed in the literature~\cite{Cimini2018a}, inspired by the derivation of Statistical Physics from Information Theory by Jaynes~\cite{Jaynes1957}. The main idea is to first create an \emph{ensemble} of all graphs having the same number of nodes as in real systems. We can then define the Shannon entropy associated with the ensemble: in order to have a maximally random benchmark, we maximize the Shannon entropy, constraining some defining quantities about the system. In this sense, by comparing the real network with our null model, all observations that cannot be explained by the constraints can be captured. Constraints can be global, as the total number of links, or local, as the degree sequence, i.e. the number of connections per node. 

In the following, we will first introduce the Bipartite Configuration Model (\emph{BiCM},~\cite{Saracco2015}), i.e. the application of the procedure described above to bipartite networks in which the degree sequences are the constraints. Then we will describe the validation procedure for co-occurrences, proposed in Ref.~\cite{Saracco2017Inferring}. Both the BiCM and the validation procedure used in the present manuscript were performed using the {\tt bicm}\footnote{\url{https://pypi.org/project/bicm/}} python module included in {\tt NEMtropy}\footnote{\url{https://pypi.org/project/NEMtropy/}}; the methods used to solve BiCM system of equations implemented in {\tt NEMtropy} and {\tt bicm} can be found in Ref.~\cite{Vallarano2021}.

\subsubsection{Formalism}
In a bipartite network, nodes are divided into two sets, called \emph{layers} and links exist only between nodes belonging to different layers. Given a bipartite network $G_\text{Bi}$, let us call its layers $\top$ and $\bot$, respectively, and $N_\top$ and $N_\bot$ their dimensions. Then, a bipartite binary network is completely described by its biadjacency matrix $\mathbf{B}$, i.e. a $N_\top\times N_\bot$ rectangular matrix whose generic entry $b_{i\alpha}$ is either 1 or 0 if there exists a link connecting node $i\in\top$ and $\alpha\in\bot$ or not. The degree of a generic node $i\in\top$ ($\alpha\in\bot$) is simply $k_i=\sum_{\alpha\in\bot}b_{i\alpha}$ ($h_\alpha=\sum_{i\in\top}b_{i\alpha}$). In the following, quantities related to real networks will be indicated with an asterisk $*$.

\subsubsection{BiCM}
Let us call $\mathcal{G}_\text{Bi}$ the ensemble of graphs of the Bipartite Configuration Model in which each representative graph $G_\text{Bi}\in\mathcal{G}_\text{Bi}$ is a $N_\top^*\times N_\bot^*$ bipartite network\footnote{Since all networks will have the same number of nodes in each layer, asterisks will fall only from $N_\top^*$ and $N_\bot^*$.}. We define the Shannon entropy associated with the system as
\begin{equation}\label{eq:S}
    S=-\sum_{G_\text{Bi}\in\mathcal{G}_\text{Bi}} P(G_\text{Bi})\ln P(G_\text{Bi}).
\end{equation}

We can perform a constrained maximization of the Shannon entropy using the methods of Lagrangian multipliers, the constraints being the degree sequences of both layers, i.e. $k_i,\,\forall i\in \top,$ and $h_\alpha,\,\forall\alpha\in\bot$. In this way, we will achieve a benchmark that is maximally random, but, in which the average degree sequences are equal to the ones observed in the real system. Therefore, by observing deviations from the null model we will detect all structures of the real system that cannot be simply explained by the constraints. Such a procedure can be achieved through the maximization of the function $S'$ defined as
\begin{equation*}
    S'=S+\beta \Big(1-\sum_{G_\text{Bi}\in\mathcal{G}_\text{Bi}}P(G_\text{Bi})\Big)+\sum_{i\in\top}\theta_i\Big(k_i-\langle k_i\rangle\Big)+\sum_{\alpha\in\bot}\eta_\alpha\Big(h_\alpha-\langle h_\alpha\rangle\Big),
\end{equation*}
where $S$ is the Shannon entropy defined in Eq.~\ref{eq:S} and $\beta$, $\theta_i$ and $\eta_\alpha$ are the Lagrangian multipliers associated, respectively, to the normalization of the probability, to the degree sequence on layer $\top$ and to the degree sequence on layer $\bot$.
The maximization of $S'$ returns a probability per graph that can be written in terms of independent probabilities per link~\cite{Park2004}:
\begin{equation}\label{eq:P_G}
    P(G_\text{Bi})=\prod_{i,\alpha}\dfrac{e^{-(\theta_i+\eta_\alpha)b_{i\alpha}}}{1+e^{-(\theta_i+\eta_\alpha)}}=\prod_{i,\alpha}p_{i\alpha}^{b_{i\alpha}}(1-p_{i\alpha})^{1-b_{i\alpha}}.
\end{equation}
Eq.~\ref{eq:P_G} is just formal since we do not know the numerical value of Lagrangian multipliers $\theta_i$ and $\eta_\alpha$. This can be obtained through the maximization of the likelihood of observing the real system~\cite{Garlaschelli2008,Squartini2011a}. It can be shown that maximizing the likelihood is equivalent to set:
\begin{equation*}
    \left\{
    \begin{array}{l}
    \langle k_i\rangle=k_i^*\\
    \langle h_\alpha\rangle=h_\alpha^*.
    \end{array}
    \right.
\end{equation*}
In Section \mape{6} of the Supplementary Information, the interested reader can find a detailed description of how to use the Bipartite Configuration Model as a statistical benchmark to validate the co-occurrences observed in the real network.

\subsection{Discursive communities}
As stated above and described in detail in Section \mape{6} of Supplementary Information, the BiCM described above can be used as a statistical benchmark to highlight groups of users contributing to the formation of the same discourse. On Twitter, this translates to groups of users endorsing similar content. In Ref.~\cite{Becatti2019d} a procedure was proposed, later refined in Ref.s~\cite{Caldarelli2020b,Radicioni2021a,Mattei2022}. The rationale is to consider who are the creators of content and how to capture similarities among them. It has been observed in several studies that verified accounts, i.e. the ones for which the Twitter platform checked the identity of their owners --at least in the pre-Musk era-- are strong creators of content~\cite{Becatti2019d,Caldarelli2020b}. It is possible, then, to infer how similar they are perceived by the ``general'' public of unverified users by using a bipartite representation: if verified and unverified users are the two layers of a bipartite network in which the (undirected) links represent retweets\footnote{\mape{In this representation, a link connecting a verified users $v$ with a non verified $u$ is present if at least one of them has retweeted the other at least once. Due to the strong production activity of verified accounts, nearly all of the links are from the layer of verified users to the opposite one.}}, we can validate the projection on the layer of verified users. In this way, we will detect non-trivial similarities in the common audience of unverified users: otherwise stated, if a couple of verified users are retweeted by the same (non-verified) users, they are probably sharing similar positions. 
In the monopartite validated projection of verified users, communities were detected using an optimized version of Louvain algorithm~\cite{Blondel2008}: since Louvain is known to be node-order dependent~\cite{Fortunato2010}, the order of the nodes is shuffled 1000 and the configuration displaying the greatest value of the modularity is chosen.
The labels of the communities found through the community detection are then propagated in the retweet network: in fact, it is an old result that Twitter users endorse content created by others much more with retweets than with mentions~\cite{Conover2011,Conover2011a,Conover2012}. Since in many cases, some strong creators of content are not verified (and therefore run the risk of not getting a label), we run the label propagation algorithm on the undirected version of the retweet network: the rationale is that not only the sources give an indication of the user orientation, but also her audience. Otherwise stated, if the majority of a user audience has a clear orientation, it is presumable that the considered user also has the same one. For propagating labels, we implemented the procedure proposed in Ref.~\cite{Raghavan2007a}. \mape{This algorithm assigns the unverified user the label associated with the majority of its neighbors in the retweet network. If an unverified user has no verified users as direct neighbors, it will be assigned the label associated with the majority of unverified neighbors that have already been labeled. This continues iteratively until it converges.

\mape{The interested reader can find in Section \mape{7} a comparison between DiCos obtained using validated or non-validated projections. In summary, it has been observed that politicians are particularly clustered in the validated network~\cite{Becatti2019d,Caldarelli2020b,Caldarelli2021,Mattei2021,Radicioni2021a,Radicioni2021b,Mattei2022,Bruno2022}. Therefore, detecting community therein is particularly efficient in finding discursive communities about political subjects. On the contrary, the discursive communities calculated on the non-validated projection are much noisier.}}

\subsection{News engagement communities}

\subsubsection{URL manipulation}
To detect similarities in users' endorsement of pieces of news, we first need to pre-process the URLs contained in the various tweets. Sharing a compact version of a URL allows for the sharing of long URLs in tweets while maintaining the maximum character limit. For our analyses, we translated all shortened links into their original long versions. This enabled us to (i) read the top-level domain of the news source to assign a nutrition label using NewsGuard and (ii) use the long links as unique identifiers for each shared news item in our network models.

\subsubsection{NEC communities}
In order to find users sharing similar ``information diets'', i.e. engaging with the same URLs,  we used the same approach as in Ref.~\cite{GuarinoPGC21}. We first represented users sharing (either via tweets or retweets) URLs as a bipartite network of users and URLs. Then, we projected the information contained therein on the layer of users and finally validated the projection using the procedure described above. As mentioned in Section~\ref{sec:necs}, the fraction of validated nodes, in this case, is extremely limited, i.e. nearly $1.71\%$, signaling that most of the users' endorsement to URLs (and so pieces of news) is compatible with the random noise. Again, in order to find communities of users in the validated projection network, the reshuffled version of Louvain was used. 
% While in the context of discursive communities, the validated projection on the layer of unverified users is of limited benefit (since there are many unverified users not interacting with verified accounts), in the case of news engagement the validated projection on the layer of URLs permits to characterize more in detail the way the users make use of the news. In this case, the percentage of validated nodes is much greater, reaching $22.3\%$. Again, we implemented the Louvain algorithm to highlight communities of URLs. 

\section{Acknowledgments}
The authors thank Emanuele Brugnoli, Stefano Guarino, Walter Quattrociocchi, and Fabiana Zollo for insightful discussions.

%\section{Supplementary Material}
%Supplementary material is available at PNAS Nexus online.

\section{Funding}
Work partially supported by project SERICS (PE00000014) under the NRRP MUR program funded by the EU - NGEU; by the Integrated Activity Project TOFFEe (TOols for Fighting FakEs) \url{https://toffee.imtlucca.it/}; by the IIT-CNR funded Project re-DESIRE (DissEmination of ScIentific REsults 2.0).

\section{Author contributions statement}
All the authors designed the research, wrote and reviewed the manuscript. M. Pr. performed the research and analyzed data. 

\section{Previous presentation}
Some of the results were presented at NetSci 2023, Wien, on July 14th, 2023.

%\section{Preprints}
%An earlier version of this work is available at
%\url{https://doi.org/10.48550/arXiv.2308.01750}.

\section{Data availability}
The Twitter dataset is available from the corresponding author upon reasonable request. The association between the trustworthiness labels and the news sources that supports the findings of this study is proprietary Newsguard data.

\bibliographystyle{plain}
%\bibliography{reference}

\newpage
\appendix

\section{Literature Review}\label{sec:LiteratureReview}
The detection of echo chambers has been generally approached by the literature starting from online content whose nature is known a priori. 
Through the analysis of the social accounts that interact with specific content, e.g. via likes, shares, retweets, and comments, it has been shown how information relating to specific narratives attracts distinct communities.   %
Work in~\cite{delvicario2016}, by Del Vicario {\it et al.}, focuses on public Facebook pages divided into two groups: conspiracy theories  and news about science (conspiracy theories are `the pages that disseminate alternative, controversial information, often lacking supporting evidence'~\cite{delvicario2016}). The findings are that users are divided into homogeneous clusters: by analysing the accounts that share news about science and conspiracies, they are bound by ties of friendship in the network. Quoting the authors: `different contents generate different echo chambers, characterized by the high level of homogeneity inside them'.\\

Homogeneity is not only about friendship, but also about emotional approach and reaction to debunking attempts. 
Zollo {\it et al.}, in~\cite{zollo2016Emotional}, establish how users polarised on conspiracies express more negative feelings in their comments than users polarised on science news. 
Work in~\cite{Zollo2017Debunking} confirms how the echo chamber paradigm goes hand in hand with the confirmation bias phenomenon --the users' tendency to look for, prefer and interpret information in line with their thoughts~\cite{nickerson1998confirmation,KLAYMAN1995385},  while ignoring or downplaying evidence that contradicts their beliefs: interactions with debunking posts (i.e., posts that provide fact-checked information to specific topics) are overwhelmingly from users biased towards science or non-biased users.\\
The above examples show how echo chambers emerge by analysing thematic pages and noting that users divide into distinct communities according to the page topic. Going deeper, it also emerges that consecutively sharing users are linked by friendship links on the network.\\ 

Interestingly for the purpose of this article, other studies have instead analysed the dynamics of information exposure by considering the news URLs present in the posts. This is the case, e.g., of work by Weaver {\it et al.}~\cite{WEAVER201918}, in which the network of densely-connected news articles is constructed. It starts from the number of news URLs shared by each user, to arrive at the weighted network of news URLs in which the weights between two URLs identify how many users have re-shared the URL pair. 
Leveraging a state-of-the-art community detection algorithm, communities of co-shared news items are found, distinct in terms of political leaning (i.e., left-leaning and right-leaning).\\ 
% Leveraging a technique proposed by one of the authors of this article, 
Guarino {\it et al.}, in~\cite{GuarinoPGC21}, consider public Facebook pages, without however knowing \emph{a priori} the kind/quality/reputability of their content. Focusing on the activity of users sharing links to pieces of online news, the authors construct the bipartite network of users/shared URLs\ and apply the Bipartite Configuration Model (BiCM) introduced in~\cite{Saracco2017Inferring} to project the bipartite network on the two levels, the user level and the URL level. Applying the BiCM assures that two accounts (resp., two URLs) are connected if the number of URLs shared by both the accounts (resp., if the number of accounts sharing both the URLs) is so large that it cannot be explained by the degree distribution of the two layers only. %The analysis in~\cite{GuarinoPGC21} is similar to that in~\cite{WEAVER201918}, but now the network of urls is not weighted, the weights being the number of accounts that shared two url. In fact, the network of urls is instead compared to an entropy-based benchmark, called BiCM: if the weights are statistically significant, the co-occurrences of news urls are validated.\\
% In this case, a  community detection algorithm reveals urls' communities pointing at propaganda and disinformation articles and users' communities pushing  this type of news.\\

% Importanza di benchmark statistico
%Beside Ref.~\cite{GuarinoPGC21}, De Clerck {\it et al.} stress the importance of using proper statistical benchmarks for the analyses of Online Social Networks~\cite{declerck2022a,declerck2022b}: in fact, such systems are affected by strong noise and detecting genuine signals is fundamental in order to drive the proper conclusions. Recently, a stream of research focused on the application of maximum-entropy null models to different kind of Twitter datasets, to analyse {\it discursive communities}~\cite{Becatti2019d,Caldarelli2021,Radicioni2021a, Mattei2022}, non trivial coordinated activities of social bots~\cite{Caldarelli2020b,Bruno2022}, the effective flow of disinformation~\cite{Caldarelli2021} and the semantic networks of various debates~\cite{Radicioni2021a,Radicioni2021b,Mattei2021}. Among those, discursive communities are particularly relevant for the aim of the present manuscript. Having Twitter as the reference platform, discursive communities are groups of users  that interact among themselves via retweet (i.e., sharing content created by others) and contributing the same discourse~\cite{Radicioni2021a}. Mattei {\it et al.} in~\cite{Mattei2022} showed that discursive communities consist of homogeneous clusters, in case the argument of the online debate is political or societal.\\

\section{Keywords for data collection}
\begin{table*}[h!]
\centering
\begin{tabular}{l|l}
    \hline
   \bf{Keywords} & \bf{English meaning}\\
    \hline
    vax, vaccino, vaccini, vaccinarsi &Variants of the word `vaccination' \\
   novax  & A person against vaccination\\
    Astrazeneca, Pfizer-BioNTech, Moderna, Sputnik&  Covid-19 vaccines\\
    greenpass&The certificate of vaccination\\& or of recover from the disease\\
    \hline
\end{tabular}
\caption{\textbf{Keywords used for collecting tweets about the Twitter debate on the Covid-19 vaccination campaign.} Keywords have been searched in Italian, English meanings on the right. \label{tab:keywords}}
\end{table*}

\section{News Engagement Communities of URLs}\label{sec:NECURL}
\mape{Similar to what was done in the main text to find validated communities of users, it is possible to analyze the ties between users and the URLs present in their tweets and retweets to find validated communities of URLs. 
Again, the procedure involves a comparison between the observations and an entropy-based benchmark:} 
% if two users tweeted (or retweeted)  the same URLs significantly more than the benchmark, we conclude that the two users share the same information diet in a statistically significant way. 
if two URLs appear in the tweets (or retweets) of the same users significantly more than the benchmark, these URLs pass the validation procedure: 
We can thus 
 identify groups of URLs shared by the same users. 
URL communities 
% and URL communities 
 that pass the validation are called  %respectively, 
 \emph{news engagement communities} of URLs, 
% and URLs, As done in the main text for user communities, here we characterize the news engagement communities of URLs,  aka \urlnec s that result after applying the validation procedure to the original dataset. 

Table~\ref{tab:URL_community_not_clustered} summarizes 
the breakdown of URLs into \urlnec s:
Only  22\% of all URLs are validated by our procedure.

% shows that approximately $22\%$ of the traffic (considering URLs alone) can be associated with clustered URLs, i.e., belong to a \urlnec. 

\begin{table}[ht]
\centering
%\begin{center}
\caption{\textbf{URLs in  \urlnec .} Although validated URLs represent a limited minority of all URLs in the dataset, their percentage is greater than their user analogues  (i.e. 22\% vs. 2\%). 
\label{tab:URL_community_not_clustered}}

\begin{tabular}{cc}
\hline
Comm. ID &  No. URL \\%No. Users &  verified &    \\%& T &  N \\
\hline
Non-validated &  179,175 \\%& 31219 &       435 &  \\%&    39.0 &     6.0 \\
         
       % 4 &   7422 &        16 & 38731 &    30.0 &    49.0 \\
       %   1 &    674 &         1 &  2234 &     0.0 &    84.0 \\
       %   6 &    876 &         0 &  2019 &     3.0 &    95.0 \\
       %  11 &   1064 &         4 &  1681 &    93.0 &     0.0 \\
       %   7 &    521 &         0 &  1613 &     0.0 &    81.0 \\
       %  10 &     65 &         1 &  1557 &     0.0 &    30.0 \\
       %   9 &    584 &         1 &  1238 &     0.0 &   100.0 \\
       %   5 &    311 &         0 &   562 &     4.0 &    95.0 \\
       %  12 &    175 &         1 &   408 &    10.0 &    84.0 \\
       %   3 &    101 &         1 &   393 &     0.0 &   100.0 \\
       %   0 &    161 &         0 &   365 &     0.0 &   100.0 \\
       %   8 &    149 &         0 &   308 &     0.0 &   100.0 \\
       %  13 &    253 &         0 &   304 &     0.0 &    82.0 \\
       %  14 &     42 &         0 &    59 &     0.0 &    63.0 \\
        %  2 &     23 &         0 &    32 &     0.0 &    88.0 \\
        
Validated & 51,504 \\%& 25 & 51504 \\%&  &  & \\        

\hline
\end{tabular}
%\end{center}
\end{table}

%La Figure~\ref{fig:validated_URL_network} mostra le \urlnec s emergenti dai dati. Come accennato una \urlnec \ rappresenta una community di URLs connessi tra loro tramite una relazione di similarità. In Figure~\ref{fig:validated_URL_network} oltre alla divisione nelle varie \urlnec \ per ogni nodo viene mostrato anche la nutrition label associata alla sorgente che pubblica il contenuto (suggerita da NewsGuard).

\begin{table}[ht]
%\begin{center}
\centering
\caption{\textbf{Statistics for \urlnec s.} While community 4 is by far the largest, there are 6 other communities with more than 1000 URLs. 
% in this sense, it is not surprising to observe such homogeneity in the trustworthiness of the relative publishers. 
\label{tab:urlnec_reliability}}
\begin{tabular}{cccccc}
\hline
Comm. ID &  No. Users &  verified &  distinct URL &  sources &   No.URL \\
\hline
        % -1 &  31219 &       435 &          64934 &     5758 & 179175 \\
          4 &   7422 &        16 &            223 &       71 &  38731 \\
          1 &    674 &         1 &             79 &        9 &   2234 \\
          6 &    876 &         0 &             87 &        5 &   2019 \\
         11 &   1064 &         4 &             21 &        4 &   1681 \\
          7 &    521 &         0 &             64 &        6 &   1613 \\
         10 &     65 &         1 &             27 &        9 &   1557 \\
          9 &    584 &         1 &             58 &        1 &   1238 \\
          5 &    311 &         0 &             23 &        6 &    562 \\
         12 &    175 &         1 &             28 &        4 &    408 \\
          3 &    101 &         1 &             79 &        1 &    393 \\
          0 &    161 &         0 &             55 &        1 &    365 \\
          8 &    149 &         0 &             25 &        1 &    308 \\
         13 &    253 &         0 &              3 &        3 &    304 \\
         14 &     42 &         0 &              4 &        2 &     59 \\
          2 &     23 &         0 &              3 &        3 &     32 \\
\hline
\end{tabular}
%\end{center}
\end{table}

More details can be found in Table~\ref{tab:urlnec_reliability}, which shows some information about the different \urlnec s. 
\urlnec \ 4 is the largest in terms of both size (consisting of 223 nodes) and impact on the overall dataset, as measured by the number of shares ($\sim39$k shares). 
The remaining \urlnec s can be distinguished based on the order of magnitude of the shares: we have 6 communities whose URLs were shared thousands of times, and other communities whose URLs were shared hundreds or dozens of times. Remarkably, in all but 4 of the \urlnec s, the number of different sources is quite limited (where source means the online news outlet that published the news to which the URL points).\\%In a sense, \urlnec s are capturing some niche of attention of the users.\\

To get a finer description of \urlnec s, we examine the frequency of untrustworthy news sources in them. 
For each URL pointing to a news article, we consider the corresponding second-level domain\footnote{\url{https://en.wikipedia.org/wiki/Domain_name}}, which refers to the name directly to the left of .com, .net, and other top-level domains (such as \url{nytimes.com} and \url{latimes.com}). We then associate the domains with the publishers, annotating the former with the reputation labels provided for the latter by the NewsGuard site (\url{https://www.newsguardtech.com/}).
In this sense, the trustworthiness of a URL is inherited from the trustworthiness of its domain/publisher, i.e. a news item is considered more or less trustworthy depending on the trustworthiness of its publisher.
According to the NewsGuard classification, the labels T (`Trustworthy'), N (`Not trustworthy') and UNC (`Unclassified') stand for the level of trustworthiness of the publisher. For more details on how the information is processed by NewsGuard, see Section~\ref{sec:news_source_reliability}. %The labels were obtained from NewsGuard (\url{https://www.newsguardtech.com/}).\\

\begin{figure}[h!]
\begin{center}
\includegraphics[width=.7\textwidth]{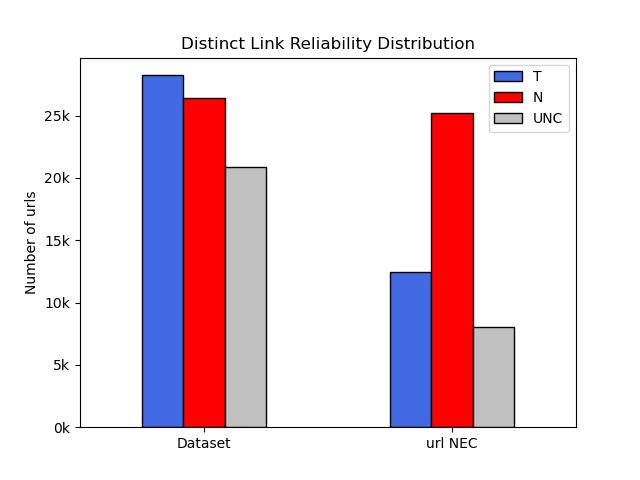}\hfill
\end{center}
\caption{\textbf{Number of distinct URLs pointing to  publishers tagged as `Trustworthy' (T), `Not trustworthy' (N), or `Unclassified' (UNC).}  \urlnec 
 s capture almost all non-trusted unique URLs in our dataset.
\label{fig:frequence_T_N_url_nec}}

\end{figure}

The first observation is that \urlnec s are a receptacle of untrustworthy sources, see Fig.~\ref{fig:frequence_T_N_url_nec}. With respect to the total number of distinct URLs in our dataset, \urlnec s capture less than half of the trustworthy ones, but almost all of the untrustworthy ones. %More than one-half of URLs in \urlnec s come from not trustworthy sources.

\begin{figure}[h!]
\begin{center}
\includegraphics[width=.70\textwidth]{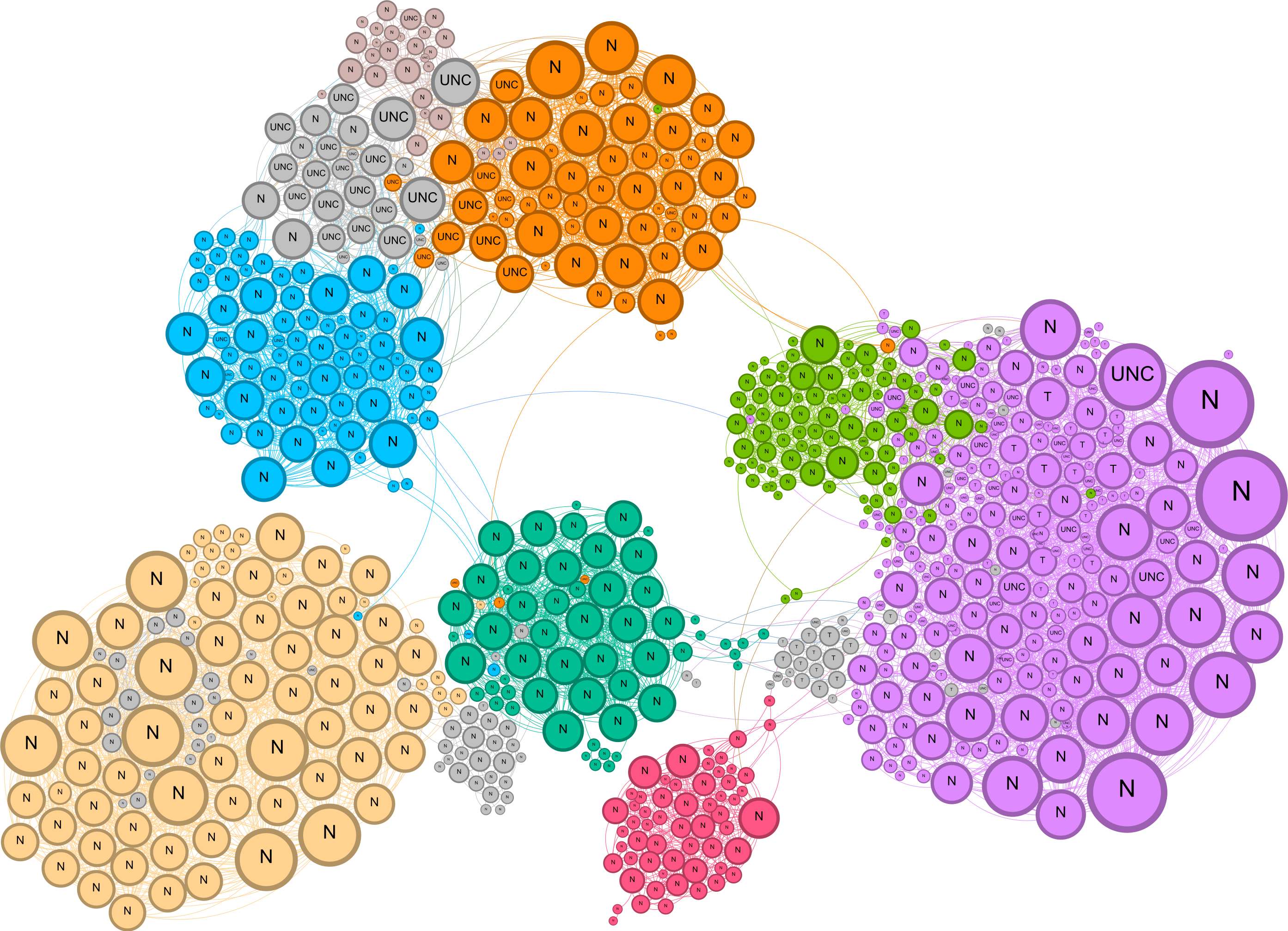}\hfill
\end{center}
\caption{\textbf{Network representation of \urlnec s.} The labels on the nodes represent the trustworthiness of the domain of the URL as labeled by NewsGuard (T for `Trustworthy', N for `Not trustworthy', UNC for `Unclassified' sources). Each community shows a strong homogeneity in the trustworthiness label. \label{fig:validated_URL_network}}
\end{figure}

Fig.~\ref{fig:validated_URL_network} pictorially shows the network of \urlnec s, as it emerges from the data. 
The different communities show a strong homogeneity in the trustworthiness of their sources.

%\paragraph{URL NECs and reputability of the news sources}

To investigate more deeply the level of homogeneity of the single community in terms of the trustworthiness label of URLs within them, we consider the frequency of trustworthy and untrustworthy sources of URLs therein. For the $i$-community of \urlnec s, if $R$ is the trustworthiness value (either $T$ or $N$), we define $purity_{\small{R}} (\text{URL NEC}_i)$ the frequency of URLs from $R$ domains, i.e. 
\begin{equation}\label{eq:purity}
purity_{\small{R}} (\text{URL NEC}_i) = \frac{ | U_{i}^{\small{R}} | } { | U_{i} | },  
\end{equation}
where $U_{i} = \{URL_{1}, \dots, URL_{\small{n}} \}$ is the set of all the URLs in the $i$-community and $U_{i}^{\small{R}} \subseteq U_{i}$ is the subset of $U_{i}$ that contains only URLs with  trustworthiness $R$. The purity defined in Eq.~\ref{eq:purity} can be interpreted as the probability of extracting an R-reputable URL in the $i$-th \urlnec.
If $m$ is the number of different \urlnec s, we can define $purity_{\small{R}}(\cup_i\text{URL NEC}_i)$ as the frequency of URLs from $R$ domains in all \urlnec s:
\begin{equation}\label{eq:overall_purity}
purity_{\small{R}}(\cup_i\text{URL NEC}_i)= \dfrac{\sum_{i=1}^m |U_i^{\small{R}}|}{\sum_{i=1}^m|U_i|}
\end{equation}

\begin{comment}
\begin{equation}\label{eq:overall_purity}
overall\_purity^{\small{R}} = \sum_{i=1}^{ \big\| NECs \big\| } purity_{i}^{\small{R}}
\end{equation}
\end{comment}
%The $overall\_purity^{\small{R}}$ expresses in a single numerical value the degree of homogeneity of all the $NECs$ of a single type with respect to the reliability value R.
To have a benchmark for the purity of \urlnec s, we also consider a purity measure for URLs that do not belong to any community: 
\begin{equation}\label{eq:unclust_purity}
purity_{\small{R}} (\overline{\cup_i\text{URL NEC}_i}) = \frac{ | U_{-1}^{\small{R}} | } { | U_{-1}| },  
\end{equation}
where the set of URLs that do not belong to any community is denoted as $U_{-1}$.\\
%Comparing the values of $overall\_purity^{\small{R}}$ and $purity_{-1}^{\small{R}}$ provides insight into the characteristics of news articles associated with nodes that belong or not to a community. We can also compare $overall\_purity^{\small{R}}$ and $purity_{-1}^{\small{R}}$ with $purity_{i}^{\small{R}}$, to evaluate how closely each community is aligned with the global measures.
%Comparing the values of $purity^{\small{R}}_\text{NECs}$ and $purity_{-1}^{\small{R}}$ provides insight into the characteristics of news articles associated with nodes that belong or not to a community. We can also compare $purity^{\small{R}}_\text{NECs}$ and $purity_{-1}^{\small{R}}$ with $purity_{i}^{\small{R}}$, to evaluate how closely each community is aligned with the global measures.
\begin{figure}[h!]\begin{center}
\includegraphics[width=.90\textwidth]{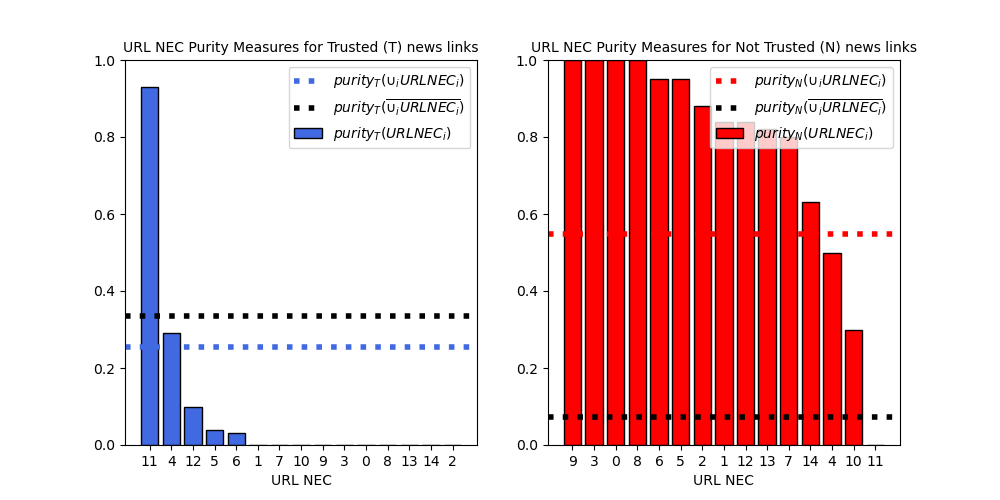}\hfill
\caption{\textbf{Purity levels of \urlnec s.} On the left trustworthy URLs, on the right untrustworthy ones. While the T purities of the individual communities are particularly low (less than 0.2 in most cases), the analog N purities are greater than $0.6$ for most of the \urlnec s.
\label{img:URL_per_comm_purity_T_N}}
\end{center}
\end{figure}

Fig.~\ref{img:URL_per_comm_purity_T_N} shows the homogeneity of \urlnec\ communities concerning trustworthy (\emph{T}, left panel) and untrustworthy (\emph{N}, right panel) news sources. On the x-axis there are the \urlnec\ communities, denoted by their ids, while the y-axis reports the purity of each community. 
The blue dotted line indicates  $purity_{\small{R}}(\cup_i\text{URL NEC}_i)$, the black dotted line indicates $purity_{\small{R}}(\overline{\cup_i\text{URL NEC}_i})$.\\
Focusing on the $purity_{\small{R}}(\cup_i\text{URL NEC}_i)$ lines, on average \urlnec s have higher N purity URLs ($\sim0.548$) compared to T URLs ($0.254$). 
Such a result suggests that URLs belonging to \urlnec s represent niches of misinformation sources, and it is corroborated by the observation that most of \urlnec s refer to a limited number of different sources (see Table~\ref{tab:urlnec_reliability}).

\section{Article's reputability measure (NewsGuard)}\label{sec:news_source_reliability}
One of the aims of the work is to characterize the variety of domains circulating within the dataset, both in terms of type (e.g., news site, marketplace, social platform, etc.) and transparency and credibility (only in the case of news sites). 
% Here, we report a series of analyses related to the domains that mostly appear in the tweets of the validated network of verified users. \final
In this paper, we refer to domains as the  `second-level domain' names\footnote{\url{https://en.wikipedia.org/wiki/Domain_name}}, i.e., the names directly to the left of .com, .net, and any other top-level domains. For instance, we consider domains \url{nytimes.com}, \url{guardian.com}, \url{corriere.it}.  

The domains have been tagged according to their degree of credibility and transparency,  as indicated by fact-checking website NewsGuard (\url{https://www.newsguardtech.com/)}. 
 The NewsGuard initiative was born from the joint effort of journalists and software developers, aiming at evaluating news sites according to  criteria concerning credibility and transparency. For evaluating the credibility level of a source of information, NewsGuard metrics consider, e.g., whether the news source regularly publishes false news, whether it distinguishes between facts and opinions, or whether it does not correct a wrongly reported news. For transparency, instead, NewsGuard evaluation takes into account, e.g., whether owners, founders or authors of the news source are publicly known, or whether advertisements are easily recognizable\footnote{Details on the news site evaluation starting from the estimate of the assessment criteria are available at: \url{https://www.newsguardtech.com/ratings/rating-process-criteria/}.}.
% After combining the individual scores obtained out of the nine criteria, NewsGuard associates to a news source a global score from 1 to 100, where 60 is the minimum score for the source to be considered reliable.  When reporting the results, the toolkit  provides details about the criteria which passed the test and those that did not.
%The details of this procedure are reported below.

% As a first step, we considered the network of verified accounts, whose communities and subcommunities have been shown
% in Figure~\ref{fig:subcomm_netwk}. 
%In our dataset, we labelled all domains that had been shared at least 20 times in tweets and retweets. 

%\input{tab_SI_5.tex}
\begin{table}[ht!]
\centering
\begin{tabular}{c|l}
\hline
label & \text{description}\\
\hline
T & Trustworthy news source\\
%$\sim\text{R}$ & Quasi Reputable news source\\
N & Not trustworthy news source\\
P & Platform (e.g., reddit.com, twitter.com, facebook.com)\\
S & Satire\\
UNC & Unclassified source\\
\hline
\end{tabular}

\smallskip
\caption{Tags for domain labelling. Tags are inherited from NewsGuard. The UNC tag indicates that NewsGuard did not tag that domain. 
\label{table:domains-tags}}
\end{table}

Table~\ref{table:domains-tags} shows the tags associated with domains. In the manuscript we shall be interested in quantifying the reliability of news sources that were publishing during the period of interest. 
Thus, we will not consider those sources corresponding  to  social networks (tag P). Also, we will not consider satiric news (tag S). 
% \fasa{; nevertheless, the information regarding their frequency are available for the interested readers in the Supplementary Material.} 
Tags T  and N in Table~\ref{table:domains-tags} are used only for news sites, be they newspapers, magazines, TV or radio social channels, and they stand for `Trustworthy' and `Not trustworthy',  respectively.

\section{Exposure of users to misinformation in echo chambers}\label{sec:rep}

To provide a finer characterization of users' exposure to misinformation in echo chambers, we `recycle' the purity definition of Section~\ref{sec:NECURL}, with one crucial difference: there, the purity measure was applied to different sets of URLs from time to time; here, we apply it to all messages shared by different sets of users. Thus, in the present case, if a URL has been shared multiple times, we consider the repetitions.
The rationale for this is to characterize echo chambers in terms of the extent to which links to news stories from untrustworthy news publishers circulate within them.
If $|\text{EC}_i(\text{URL})|$ and $|\text{EC}_i(\text{URL}; R)|$ count, respectively, the number of messages containing a URL and a R-reputable URL shared by users in echo chamber $i$, with a little abuse of notation we can define a purity for echo chamber as
\begin{equation}\label{eq:purity_ec}
purity_{\small{R}}(\text{EC}_i) = \dfrac{ |\text{EC}_i(\text{URL}; R)|} {|\text{EC}_i(\text{URL})|}.  
\end{equation}

\begin{figure}[h!]\begin{center}
\includegraphics[width=.90\textwidth]{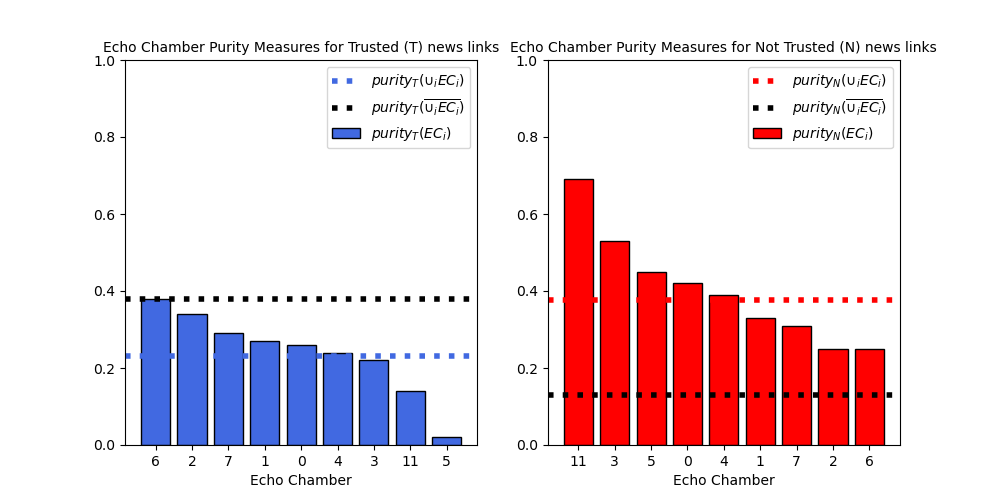}\hfill
\caption{\textbf{Purity levels of echo chambers.} On the left trustworthy URLs, on the right not trustworthy URLs. While the $purity_{\small{T}}(\overline{\cup_i\text{EC}_i})$ value is greater than its counterpart in the echo chamber, $purity_{\small{N}}(\overline{\cup_i\text{EC}_i})$ is lower than the value measured in echo chambers. 
\label{img:USR_per_comm_purity_T_N}}
\end{center}
\end{figure}

Analogously to what was done in Subsection~\ref{sec:NECURL}, we can define $purity_{\small{R}}(\cup_i\text{EC}_i)$ and $purity_{\small{R}}(\overline{\cup_i\text{EC}_i})$, respectively for all users in echo chambers and for all users outside echo chambers.
The results of the analysis is reported in Fig.~\ref{img:USR_per_comm_purity_T_N}: on the x-axis there are the echo chambers denoted by their ids, on the y-axis the purity values. 
On the left panel, purities are related to trustworthy URLs. On the right panel, purities are related to untrustworthy URLs. The blue dotted line indicates  $purity_{\small{R}}(\cup_i\text{EC}_i)$, the black dotted line indicates $purity_{\small{R}}(\overline{\cup_i\text{EC}_i})$.

% we can appreciate the difference between the values of $overall\_purity^{\small{T}}$ and $overall\_purity^{\small{N}}$: 

Focusing on the $purity_{\small{R}}(\cup_i\text{EC}_i)$ lines, echo chambers on average have a higher purity with respect to untrustworthy URLs ($\sim0.377$) compared to trustworthy ones ($\sim0.232$). In other words, when a user posts a message containing a URL in an echo chamber, the probability that it points to an untrustworthy news source is close to 0.4; for some echo chambers, this probability is even much higher than 
this. %the purity value T 
As in the case of the purity for \urlnec s, if we compare the $purity_{\small{R}}(\cup_i\text{EC}_i)$ values against $purity_{\small{R}}(\overline{\cup_i\text{EC}_i})$, there is a trend reversal in passing from T to N: the $purity_{\small{T}}(\overline{\cup_i\text{EC}_i})$ value is greater than its counterpart in the echo chamber while $purity_{\small{N}}(\overline{\cup_i\text{EC}_i})$ is lower than the value measured in echo chambers. 
This finding is worrisome because users in echo chambers are particularly polarized and committed, basing their beliefs on low-quality news. However, it is important to remember that the formation of echo chambers, while alarming in itself, is generally unrelated to the quality of news sources.

\section{Validated projection of bipartite networks}\label{sssec:val_proj}
The BiCM null model introduced in Subsection 4.1.2 of the main text can be used to validate the co-occurrence network defined from a bipartite one. Consider two nodes $i, j\in\top$: the number of co-occurrences between them is
\begin{equation}\label{eq:V}
V^{ij}=\sum_{\alpha\in\bot}V^{ij}_\alpha=\sum_{\alpha\in\bot}b_{i\alpha}b_{j\alpha}.
\end{equation} 
As mentioned in the subsection above, the probability of observing a graph $G_\text{Bi}$ is factorised in terms of probabilities of the existence of a single link. Therefore the probability that both nodes $i,j$ link a single node $\alpha \in\bot$ is simply
\begin{equation*}
    P(V^{ij}_\alpha)=p_{i\alpha}p_{j\alpha},
\end{equation*}
where $V^{ij}_\alpha$ is defined in Eq.~\ref{eq:V}.
In general, given node $i\in\top$, all $p_{i\alpha}$ are different, depending on the degree $h_\alpha$. In this sense, the BiCM probability distribution of $V^{ij}$ is the generalization of the binomial distribution in which each event $V^{ij}_\alpha$ has a different probability. Such a distribution is known in the literature with the name of Poisson-Binomial distribution. For each observed co-occurrence, we can then calculate its p-value~\cite{Saracco2017Inferring}.\\

Finally, all p-values are validated using a multiple-test hypothesis. In the present work, we use FDR~\cite{Benjamini1995}, since it permits to control the number of False Positives. In a nutshell, the FDR procedure prescribes ordering all p-value from the lowest to greatest, i.e. $\text{p-value}_1\le\text{p-value}_2\le\cdots\le\text{p-value}_n$. Then, if $n$ is the total number of tests, the effective threshold is given by the greatest $i$ satisfying
\begin{equation*}
    \text{p-value}_i\le i\dfrac{\alpha}{n},
\end{equation*}
where $\alpha$ is the statistically significant threshold. In the present analysis $\alpha=0.05$. %The entire procedure of the validated bipartite network projection is summarised in Fig.~\ref{fig:architecture}.

\section{\mape{Validated vs non-validated discursive communities}}\label{sec:DiCoSi}

\begin{figure}[ht!]
\begin{center}
\includegraphics[width=\textwidth]{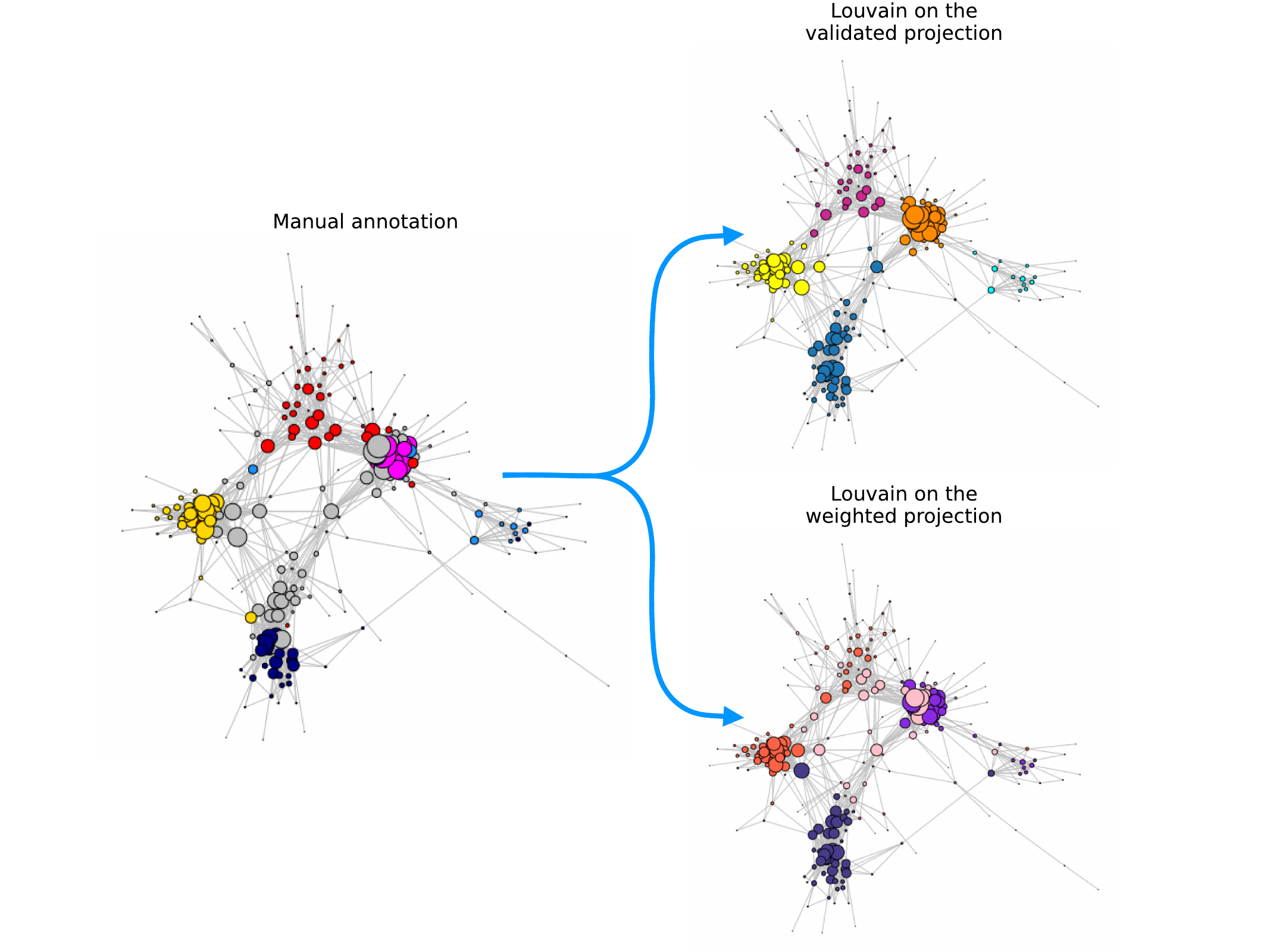}\hfill
\caption{\mape{\textbf{Comparison between the results of different community detections on the validated network of verified users.} On the left, only politicians' accounts are colored according to their political affiliation (other verified accounts are gray). The first observation is that politicians with similar orientations cluster together in the validated projection. In this sense, a community detection run on this network returns partitions that are coherent with these political clusters (top right panel; nodes with the same color belong to the same community). The same is not quite true for a community detection algorithm run on the non-validated projection: in the latter case, the partitions only partially capture the political orientations present (lower right panel; again, nodes with the same color belong to the same community).}
\label{fig:val_non_val}}
\end{center}
\end{figure}

\mape{Let us summarize the procedure for inferring the presence of discursive communities (DiCo) in our dataset, as described in the main text. Our approach focuses on the bipartite network of verified vs. unverified accounts, where a link represents the presence of at least one retweet from the unverified to the verified user. The network is then projected into the layer of verified users, resulting in a monopartite network in which the weights of the link represent the number of common (unverified) retweeters, i.e. the co-occurrences. Finally, the network is validated by comparing the empirical values with a maximum entropy null model (the BiCM~\cite{Saracco2015}), including the information of the bipartite degree sequences.

At first glance, the validation procedure may seem like an unnecessary complication. The goal of the analysis is to extract similarities in the creation of new content based on common audiences, and it can be argued that even without extracting the significant structure of the network, the standard algorithms for community detection can find the relevant network structure.\\

Before directly comparing the results in the case of our dataset, let us first provide a methodological argument in favor of using the validated projection instead of the entire projection network. As mentioned above, the output of the procedure is a monopartite network in which connections are present if the co-occurrences cannot be explained by the bipartite degree sequences. In this sense, the structure of the network is inferred by discounting the \emph{original} bipartite information. 
If, instead, the projection network is not validated, the communities in the network are inferred using the information about the projected network, i.e., some kind of information \emph{derived} from the original bipartite system. Note also that knowing the value of the co-occurrences does not allow going back to the bipartite structure of the system and causes a loss of information~\cite{Guillaume2004}. In this sense, the use of the original information available from the data should be preferred.\\ 

Nevertheless, the implications of such a choice could still be  limited in our dataset and, therefore, we will examine the results of the different approaches. The first observation, already highlighted in many papers~\cite{Becatti2019d,Caldarelli2020b,Caldarelli2021,Radicioni2021a,Radicioni2021b,Mattei2021,Mattei2022,Bruno2022}, is that, when the debate is political or societal (as in the case of our dataset), the accounts of politicians and political parties tend to cluster, according to their orientation, in the validated network of verified users. This is also the case for our dataset, as can be seen in the left panel of Fig.~\ref{fig:val_non_val}: the colored nodes represent the accounts of political parties and politicians, where the color is related to their political alliance\footnote{In dark yellow, the Movimento 5 Stelle; in dark blue, the right-wing parties Lega and Fratelli d'Italia; in sky blue, the center-right party Forza Italia; in magenta, the center-left party Italia Viva; in red, the democratic alliance, including PD (the Italian Democratic Party), +Europa, the Socialist Party, and the Green Party. In gray, other verified users whose political orientation is not given \emph{a priori}, such as journalists, media, artists, NGOs, etc.}. %Accounts of politicians sharing similar views are indeed clustered. 
The only exceptions are some Italia Viva accounts that are merged with some center-left politicians. Such behavior is justified by the fact that Italia Viva was created by politicians who left the PD because they were not satisfied with the current leadership. In this sense, it is not surprising to find links between former party members.

The Louvain algorithm, run on the validated projection, captures such groups, see the top right panel of Fig.~\ref{fig:val_non_val} (nodes displaying the same colors belong to the same community).\\

Even if running the (weighted) Louvain algorithm on the entire co-occurrence network yields, by definition, different results, they could still provide a coherent partition of the validated projection, since it represents the core of the co-occurrence network. Remarkably, discounting inferred information has a cost: the obtained partition is less coherent with the political orientations of the verified users than the former one, see the lower right panel of Fig.~\ref{fig:val_non_val}. For example, Movimento 5 Stelle and the center-left alliance are mixed. The situation is even worse for Italia Viva, which is split in 2, partly joining the center-left alliance accounts and partly mixed with Forza Italia. In this sense, we can say that the community detection on the validated projection gives cleaner partitions than those calculated on the non-validated network. Finally, comparing modularities computed on different types of networks is not particularly informative, but it can still give a rule-of-thumb idea about the organization of the network: in the case of the validated network, the modularity is $Q\simeq0.66$, while in the case of the non-validated network, it is $Q\simeq0.17$\footnote{Note that the null models implemented by the two Louvain community detection algorithms are different. On the binary validated network, it is the standard binary configuration model, which considers the information of the bipartite degree sequences. On the total co-occurrence network, it is the weighted configuration model, thus including the information of the strength sequence.}. In this sense, the validated network has a more modular structure.\\ 

In summary, in the validated projection of verified users, politicians and political parties cluster according to their political affiliation, and therefore a community detection algorithm running on the validated projection will capture these groups. Instead, a community detection algorithm running on the entire co-occurrence network of verified users, where co-occurrences is the number of common unverified retweeters, adds some noise to the partition found, and the division between opposing groups is less clean. 
}

\end{document}